\documentclass[]{aastex631}

\usepackage{graphicx}    
\usepackage{amsmath}
\usepackage{float}

\shorttitle{Helium Lines Reveal Progenitor Clues in SNRs}
\shortauthors{Das et al.}

\begin{document}

\title{Helium emission from Balmer-dominated shocks in Type Ia supernova remnants provides constraints to their progenitor systems}

\author[0000-0000-0000-0000]{Priyam Das}
\altaffiliation{E-mail: priyam.das@unsw.edu.au}
\affiliation{School of Science, The University of New South Wales, Northcott Dr., Canberra, ACT 2612, Australia}
\affiliation{Research School of Astronomy and Astrophysics, Australian National University, Canberra ACT 2611, Australia}

\author{Ivo R. Seitenzahl}
\affiliation{Research School of Astronomy and Astrophysics, Australian National University, Canberra ACT 2611, Australia}
\affiliation{Mathematical Sciences Institute, Australian National University, Canberra ACT 2611, Australia}

\author{Parviz Ghavamian}
\affiliation{Department of Physics Astronomy and Geosciences, Towson University, Towson MD 21252, USA}

\author{Ashley J. Ruiter}
\affiliation{Mathematical Sciences Institute, Australian National University, Canberra ACT 2611, Australia}

\author{J. Martin Laming}
\affiliation{Space Science Division, Naval Research Laboratory, Washington DC 20375, USA}

\author{Simon J. Murphy}
\affiliation{School of Science, The University of New South Wales, Northcott Dr., Canberra, ACT 2612, Australia}

\author{Cillian O'Donnell}
\affiliation{School of Physics, Trinity College Dublin, The University of Dublin, Dublin 2, Ireland}

\begin{abstract}
Balmer-dominated shocks in Type Ia supernova remnants offer powerful probes into collisionless shock physics and hints towards supernova progenitor environments.
Prior studies focused on the hydrogen Balmer lines, which manifest as a superposition of broad and narrow emission lines. 
Using integral-field spectroscopy with MUSE, we discovered broad and narrow helium emission lines from Balmer-dominated filaments of three Type Ia supernovae remnants in the Large Magellanic Cloud: SNR 0509-67.5, SNR 0519-69.0 and N103B. 
We detect broad and narrow He~\textsc{i} 5876~\AA~,7065~\AA\ emission in SNR 0519 and N103B and He \textsc{ii} 8236~\AA\ in SNR 0519. In SNR 0509
we detect narrow He~\textsc{i} 5015~\AA, 6678~\AA, 7065~\AA\ and 7281~\AA, with only 7065~\AA~ exhibiting a broad component.
The detection of narrow He\,\textsc{ii} challenges existing shock models, where such emission is not expected, and may indicate either incomplete ion-ion equilibration behind the shock or an origin in shock precursors.
For SNR 0509 and N103B, the neutral He/H line ratios indicate enhanced helium abundances, whereas SNR 0519 is consistent with the primordial He/H value. We therefore propose helium emission in Balmer-dominated shocks as a new diagnostic of shock physics and Type Ia supernova circumstellar environments. Although our modeling is primarily a proof of concept, it demonstrates the possibility to infer the total He-to-H abundance ratio, with dominant uncertainties arising from the assumed initial ionization fractions. Despite the uncertainties, we demonstrate that narrow helium lines can serve as effective probes of circumstellar conditions and progenitor evolution when analysed alongside reliable constraints on the preshock neutral H/He abundance ratio.
\end{abstract}


\section{Introduction}
Type Ia supernovae (SNe~Ia) are violent thermonuclear explosions of white dwarf stars in interacting binary systems. There are a number of proposed evolutionary channels and more than one explosion mechanism is thought to lead to SNe~Ia, and so their progenitors remain a matter of considerable contention, see e.g. \citet{liu2023a} and \citet{Ruiter2025} for recent reviews.

SNe Ia are commonly categorized by the mass of the exploding primary white dwarf (WD) at ignition -- e.g. near-Chandrasekhar mass (near-Ch) models and sub-Chandrasekhar mass (sub-Ch).  
Near-Ch explosions arise when a carbon-oxygen white dwarf (CO WD) approaches $\sim$1.4~M$_\odot$, either by H or He-rich accretion, or through a merger that reaches that mass \citep{Neopane2022}, leading to an ignition of a deflagration in the high‐density core that possibly transitions later to a detonation \citep{Whelan1973,Khokhlov1991, Khokhlov1997}.
Sub-Ch events involve WDs well below 1.4~M$_\odot$, whose detonation is typically triggered by a surface helium detonation, the ``double-detonation” mechanism \citep{Taam1980,Iben1987,Livne1995,Fink2007} or by merger dynamics \citep{Iben1984,Webbink1984,Pakmor2010, Pakmor2021}.  Both single and double-degenerate channels can produce either mass class, and it is the WD mass at explosion, not the donor's identity, that most directly governs the burning physics and resulting nucleosynthesis. However, an understanding of the donor's identity is a crucial consistency check of any realistic explosion scenario.

Regardless of the nature of the explosion mechanism, the explosion creates supersonically-expanding ejecta, which act as a piston pushing against the surrounding interstellar medium, giving rise to an outwardly propagating shock wave (the forward shock, or FS). This shock heats the surrounding medium, generating optical emission lines in the immediate post-shock ionization zone, as predicted by \cite{Chevalier1978} and \cite{Bychkov1979}. The forward shock is non-radiative in the initial phases of the remnant, at least until the end of the free expansion phase \citep{Truelove1999}. Spectroscopic study of the FS of supernova remnants (SNRs) with different shock speeds provides clues to the degree of electron-ion and ion-ion  equilibration, and thereby to the nature of collisionless heating in very high Mach number shocks. SNRs such as Tycho, Kepler, and the Cygnus Loop (as well as nearly a dozen more) have been observed to emit optical emission lines by collisionally-excited hydrogen Balmer lines known as Balmer-dominated shocks (BDS) \citep{tuohy1982a,smith91, Hester1994, ghavamian01, Heng2007}. 

BDS optical spectra consist of broad and narrow components as originally predicted by the theoretical work of \cite{Chevalier1980}, most prominently seen in both H$\alpha$ and H$\beta$ emission lines. Neutral hydrogen entering the shock does not `feel' the collisionless shock transition (which is maintained by plasma turbulence and electromagnetic fields).  Once immersed in the hot post-shock plasma, collisional excitation of these slow neutrals creates the narrow component hydrogen line with a line-width that reflects the pre-shock velocity distribution.  In contrast, charge exchange (transfer of an electron) between the slow neutrals and hot shock-heated protons gives rise to fast neutrals.  Subsequent collisional excitation of these fast neutral atoms gives rise to the broad components of the Balmer lines.
The full-width at half-maximum (FWHM) of the broad component is related to the shock velocity, as faster shocks create hotter ions and hence post-charge exchange neutral hydrogen atoms with a broader velocity width \citep{Chevalier1980, Raymond1991}.  However, the width of the broad H$\alpha$ rises more slowly with shock speed beyond 2000 km s$^{-1}$, due to the decline of the charge exchange cross-section at high proton energies. For a review of BDS physics, see \citet{heng2010a}.

The emission line profiles of BDS are an excellent probe of the physical state of the medium surrounding Type Ia SNRs.  BDS analysis is useful as it provides information about the composition (e.g. neutral He/H ratio) and pre-ionization conditions immediately ahead of the FS, allowing direct constraints on the nature of the accretion process leading up to the explosion \citep{woods2018,kuuttila19} as well as the degree of collisionless heating in the FS transition \citep{ghavamian01, vanadelsberg08, Morlino2012, Hovey}. The FS is also believed to be an efficient particle accelerator, creating the bulk of cosmic rays at energies below 10$^{15}$ eV \citep{Helder2009, ackermann2013}. 

Despite decades of study, optical emission from BDS has been characterized almost exclusively through hydrogen emission, leaving the role of helium largely unexplored. Helium, however, holds key diagnostic power for understanding the physics of BDS (e.g. as probes of ion-ion equilibration)  and the physical state of circumstellar material around Type Ia supernovae. In this work, we present new medium-resolution measurements of broad and narrow He lines that expand this picture across multiple Type Ia SNRs in the Large Magellanic Cloud and demonstrate the utility of helium emission in deriving neutral He/H abundance ratios in the circumstellar medium, providing new insight into these environments.

In Section 2, we review  existing reports of helium emission in BDS, outlining the observational limitations and physical interpretations discussed in the literature. We present a summary of our observations, the data reduction process, spectrum extraction, and the fitting of the emission lines in Section 3. In Section 4, we present the resulting detections and characterize the broad and narrow helium components across our target SNRs. In Section 5, we discuss the derived He/H abundance ratios and their implications for the nature of the progenitor systems. Finally in Section 6, we summarize our conclusions and highlight the broader significance of helium diagnostics in Type Ia SNRs. Appendix A describes the procedures for local background subtraction, Galactic extinction correction, and stellar continuum removal, enabling accurate extraction of emission line spectra from the MUSE datacubes. Appendix B details the statistical methods, physical parameter calculations, and error propagation used to characterize the emission lines and quantify uncertainties in the analysis.

\section{Extant Helium Detections in Balmer-Dominated Shocks}

Observations of BDS have on rare occasions revealed faint emission lines other than those of hydrogen, primarily forbidden lines attributed not to the post-shock gas but to a cosmic-ray precursor,
e.g. in Kepler's SNR \citep{blair91, Sollerman2003} and N103B \citep{Li21}.  Permitted transitions from other elements have so far been limited to those of helium in the optical, detected in SNRs such as the Cygnus Loop \citep{Ghavamian2000, raymond2015} and SN 1006 \citep{Ghavamian2002, raymond2017}. The He~\textsc{ii} 1640~\AA\ line was also detected in the Balmer-dominated shocks of SN 1006 \citep{Raymond1995, Laming1996} and the Cygnus Loop \citep{raymond2015}, providing precedent for helium diagnostics in non-radiative shocks. Aside from the single narrow He~$\textsc{i}$ 6678~\AA\ detection in SN 1006 \citep{Ghavamian2002},    broad He $\textsc{ii}$ 4686~\AA\ has been the only other optical line detected \citep{Ghavamian2002}, which unfortunately lies outside the spectral range of our MUSE data.

The very limited census of detected emission lines in BDS is likely due to their intrinsic faintness, along with observational strategies that did not target these lines. In the case of extra-galactic SNRs, interstellar reddening has hampered the detections even more, especially at bluer wavelengths.
In this work, we report the detection of helium emission lines in three young Type Ia SNRs in the Large Magellanic Cloud (LMC): SNR 0509-67.5 (hereafter SNR 0509), SNR 0519-69.0 (hereafter SNR 0519) and N103B and discuss the implications of their distinct broad and narrow components.

\section{Methods}
We have analyzed archival MUSE observations of three young extra-galactic Type Ia SNRs in the LMC obtained with the Very Large Telescope (VLT). We have identified broad and narrow hydrogen and helium emission lines from the Balmer-dominated filaments of the remnants. 

\subsection{Data acquisition}
The Type Ia SNRs used in this study were all observed using the MUSE optical integral field spectrograph \citep{bacon2010}. MUSE is a second-generation instrument mounted on the Unit Telescope 4 (UT4) of the European Southern Observatory's (ESO) VLT at Cerro Paranal. The data on SNR 0519 and N103B were obtained from  P.ID 096.D-0352[A] (P.I: Leibundgut) and SNR 0509 was obtained from P.ID 0104.D-0104(A) (PI: Seitenzahl). The combined exposure times on source for SNR 0519, N103B and SNR 0509 were $\sim$90\,min, $\sim$\,240 min, and $\sim$29\,h 15\,min, respectively. The first two targets were observed using the WFM-NOAO-N mode, without adaptive optics to compensate for atmospheric distortions.  SNR 0509 was observed using the WFM-AO mode. This mode makes use of the UT4 Adaptive Optics Facility \citep[][AOF]{Arsenault2013}, which consists of a deformable secondary mirror \citep{Arsenault2006}, the AO modules GRAAL \citep{Paufique2010} and GALACSI \cite[only the latter is relevant for MUSE operations;][]{Stuik2006}, and the 4 Laser Guide Star Facility \citep[][4LGSF]{Calia2014}, which generates artificial guide stars using four 22\,W sodium lasers. When MUSE is operating in any of its AO modes, a notch filter centered around the 589\,nm laser wavelength is placed in the light path from the source, which is evident later as a dip with NaN values in the reduced spectra \citep{Vogt2017,Vogt2023}.

\subsection{Data reduction}
We used the MUSE data reduction pipeline version 2.8.9 \citep{Weilbacher2020}, within ESOReflex version 2.11.5 \citep{freudling2013}, to perform the standard reduction procedure on the raw files. We processed each observation separately using the dedicated science frames, and the resulting datacubes were subsequently stacked to produce a deep combined cube. We performed the data reduction on Tycho, a Linux workstation with 768 GB of memory optimized for MUSE data processing at the University of New South Wales in Canberra. The standard pipeline removes sky lines using fluxes from available sky exposures, or, when unavailable, from a small fraction of the field of view. However, this automated sky line removal is not optimal for our science case because the sky lines have fluxes comparable to those of our target emission features. We therefore applied additional post-processing to the stacked datacube to further suppress sky residuals. This step included local sky subtraction, stellar continuum removal, and correction for Galactic extinction and reddening (see Appendix \ref{sec:appendixA}).

\subsection{Region Selection}
We performed spectral extraction  for each target SNR on selected regions of interest (ROIs). The ROIs were selected using \texttt{QFitsView} \citep{2012ascl.soft10019O}, which provides an interactive platform to extract regions and their spaxel coordinates. These coordinates were then used for further analysis with custom \texttt{python} routines. After the selection, integrated stellar spectra were examined to identify bright stars overlapping with the ROIs, and these regions were masked to avoid contamination from stellar emission. ROIs were chosen in the outer filaments of the remnants which exhibited strong BDS to minimize emission dominated by shock-heated ejecta. These are the sites where strong H$\alpha$, H$\beta$, and He lines are prominently detected. We chose a single region in SNR~0509 which covers the eastern limb-brightened side. The western half of SNR~0509 exhibits multiple closely spaced and overlapping filaments and we therefore avoided this hemisphere for our analysis. For SNR~0519 and N103B, ROIs were selected from three different locations, avoiding overlap of multiple filaments in both  remnants. These regions exhibited measurable differences in the centroid offsets of the broad H$\alpha$ component relative to the narrow component. Special care was taken when selecting the ROIs in N103B, as it has multiple bright, dense radiative knots that produce strong narrow forbidden emission lines. These regions are highlighted in blue in Figs.~\ref{fig:0519}, \ref{fig:N103B}, and \ref{fig:0509}.

\subsection{Line fitting with Gaussian components}
Each ROI consists of multiple spaxels, and for each region these spaxels were summed separately to extract integrated spectra. This approach increases the signal-to-noise ratio (S/N) of the faint helium lines. A custom \texttt{python} routine was used to produce each integrated spectrum, and spectral line fitting was performed via the \texttt{curve\_fit} function from the \texttt{scipy} package \citep{2020SciPy}, which employs a non-linear least-squares algorithm to optimize model parameters. To account for the broad, narrow, and intermediate line components, the fitted model function was composed of a sum of three Gaussians and a linear continuum as described by:
\begin{equation}
    \begin{aligned}
        f(\lambda) &= m\lambda + b + A_b \exp\left(-\frac{(\lambda - \mu_b)^2}{2\sigma_b^2}\right) \nonumber \\
        &\quad + A_i \exp\left(-\frac{(\lambda - \mu_i)^2}{2\sigma_i^2}\right) + A_n \exp\left(-\frac{(\lambda - \mu_n)^2}{2\sigma_n^2}\right)
    \label{eq:three_gaussian}
    \end{aligned}
\end{equation}
In the above equation, $m$ and $b$ are the slope and intercept of the linear continuum, respectively, $\lambda$ denotes the wavelength as given in the spectrum, $A$ is the amplitude of the Gaussian, $\mu$ denotes the centre of the line, and $\sigma$ denotes the standard deviation (width). Parameters of the broad, narrow, and intermediate component Gaussians are denoted by subscripts $b$, $n$, and $i$, respectively. Simpler one- or two-Gaussian models were used for lines showing fewer components. A special fit was necessary for the H$\alpha$ emission line in N103B because the remnant lies in the vicinity of a superbubble and its H$\alpha$ emission is contaminated by the [N~\textsc{ii}] 6548\,\AA/6583\,\AA\ doublet. Thus, we used a combined five-Gaussian fit to take nitrogen into account, as shown in Fig.~\ref{fig:nitrogen}. 
We tested the significance of the intermediate Gaussian components with an F-test \citep[e.g.][]{bevington2003a} and used the fitted parameters to calculate intensity, surface brightness, Doppler velocities and velocity width along with propagated errors (see~Appendix~\ref{sec:appendixB}).
The intermediate component affects the best-fit physical parameters of the broad and narrow components for the hydrogen emission lines. For this reason, we include the intermediate component in our fits, where preferred by the F-test (see Table~\ref{table5}). The presence of the intermediate component points to a significant population of fast neutrals formed through charge exchange, which can travel back upstream \citep{Morlino2012}. Recent studies have attributed the detection of this intermediate component in the north-eastern filaments of Tycho to a broad-neutral precursor. Furthermore, the intermediate component has been proposed as a diagnostic to distinguish between narrow-line broadening caused by either cosmic rays or a neutral precursor. However, an in-depth interpretation of the intermediate components lies beyond the scope of this paper.

\section{Results}
In the following sections, we present the results of our spectral analysis for the BDS in SNR 0519, N103B, and SNR 0509. As described above, for each remnant we fit the hydrogen and helium emission lines with multi-component Gaussian models to extract the broad and narrow components and to derive their corresponding velocity widths, central Doppler shifts, line intensities and ratios.

\subsection{SNR 0519}
\begin{figure*}
    \centering
    \includegraphics[width=16cm, trim={380 20 180 20}, clip]{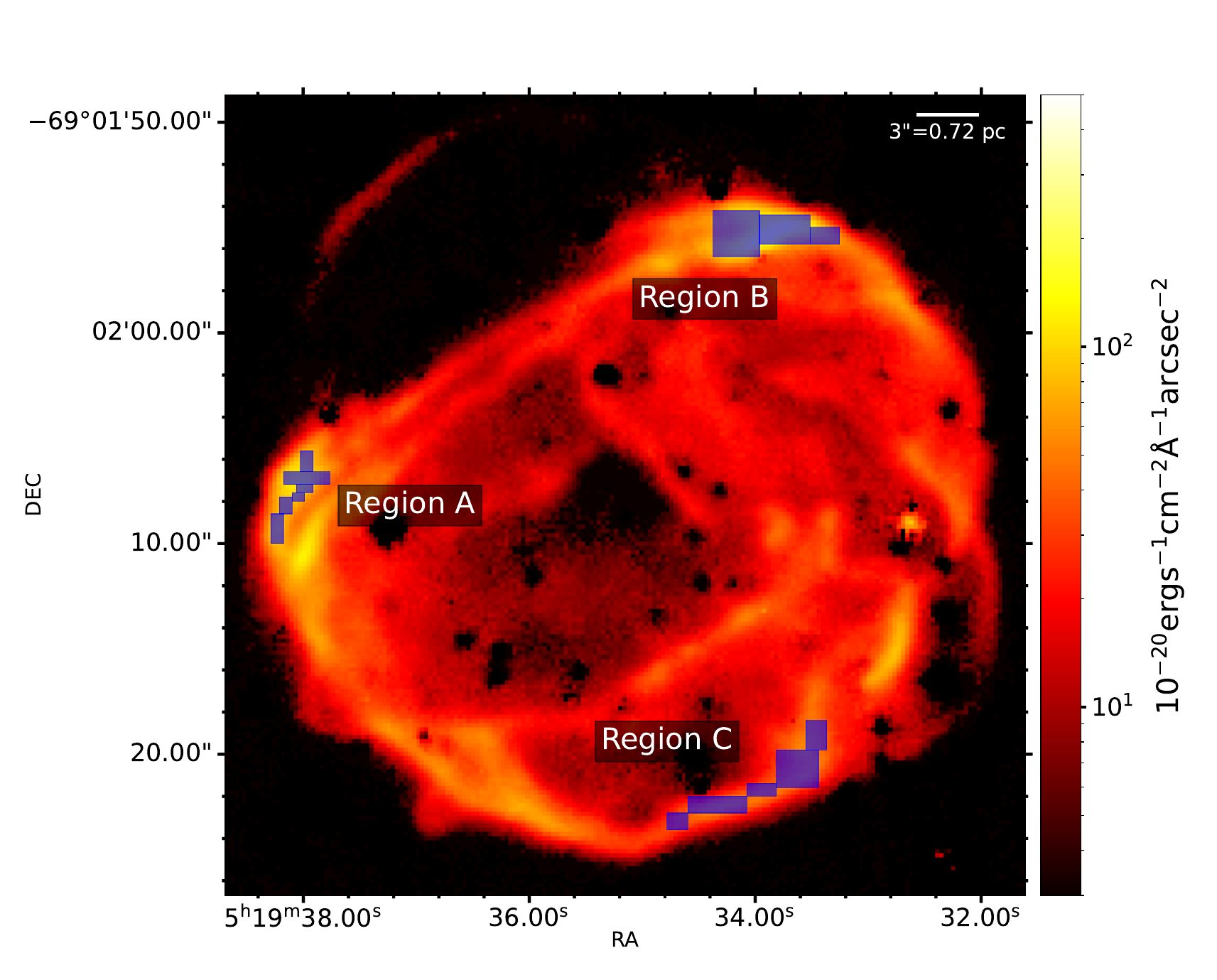}
    \caption{Continuum-subtracted mean H$\alpha$ flux density map (6473--6648\,\AA) of SNR 0519 from the MUSE datacube. A logarithmic stretch is applied for display with limits indicated on the colour bar. NaN pixels are shown in black. The blue rectangles represents the ROIs used for spectral integration and analysis.}
    \label{fig:0519}
\end{figure*}

We analyzed the Balmer-dominated spectrum of SNR 0519 using three integrated spectra from the distinct regions shown in Fig.~\ref{fig:0519}.
We detect broad and narrow helium lines in all three regions.
Specifically, we report the detection of H$\alpha$, H$\beta$, He $\textsc{i}$\,7065~\AA, He~$\textsc{i}$\,5876~\AA, He~$\textsc{i}$\,7281~\AA, and He~$\textsc{ii}$\,8263~\AA\ across all three regions. An F-test analysis on the number of components did not require consideration of an intermediate-width component, as seen, for example, in N103B \citep{Ghavamian2017}. Sample plots of the emission line profiles are shown in Fig.~\ref{fig:HeI_0519} and Fig.~\ref{fig:HeII_8236} and the best-fit parameters are given in Table~\ref{tab:table1}. The helium lines have much lower surface brightnesses (more than 100 times fainter) than the hydrogen Balmer lines. The characteristic broad and narrow line components are seen in the hydrogen line profiles, with $I_b/I_n$ observed in the range of 0.3--0.6. This is similar to values reported by \cite{Hovey}, however they observed more regions and reported a wider range of $I_b/I_n$, but the results reported here are not in conflict. The velocity width for the broad component of H$\alpha$ ranges from $\sim$1300--2400 km\,s$^{-1}$ and the broad component of H$\beta$ ranges from $\sim$1200--2500 km\,s$^{-1}$. The width of the broad component of He~\textsc{i}\,7065~\AA\ exceeds that of the hydrogen lines in Region A and, marginally, in Region B. For Region C however, which has the fastest shocks (broadest Balmer lines), broad He~\textsc{i}\,7065~\AA\ is slightly narrower than the Balmer lines.
The measured velocity widths of the broad components of He~\textsc{i} 5876~\AA\ and He~\textsc{ii}\,8236~\AA\ are in the range $\sim$300--700 km\,s$^{-1}$ and the narrow components of He~\textsc{i} and He~\textsc{ii}\,8236\,\AA\ are in the range $\sim$80--190 km\,s$^{-1}$. For the helium lines, the broad component intensity is always greater than the corresponding narrow component intensity, thus the ratio \(I_b/I_n > 1\) for helium in all regions. The observed Doppler shift for the broad component of He~\textsc{i}~5876~\AA\ and He~\textsc{ii}\,8236~\AA\  is consistent (within the margin of error) across all three regions. All the narrow He components exhibit Doppler shifts consistent with narrow H, except for narrow He~\textsc{i}~7065~\AA\ in Region C. We also note an offset of the narrow Doppler velocity of the line centroid between H$\alpha$ and H$\beta$ ($\Delta\sim$30 km\,s$^{-1}$) in all three regions.

\begin{table*}
\caption{\label{tab:table1}Physical parameters of hydrogen and helium emission lines measured in SNR~0519 after correcting for Galactic reddening only (see Appendix~\ref{sec:appendixA}). The quantity $I$ gives the relative line intensity normalized to the narrow H$\beta$ component (by convention $I_{H\beta,n}=100$); the absolute surface brightness of narrow H$\beta$ is listed in the table header. $W$ denotes the velocity full width at half maximum (FWHM) of the fitted Gaussian components and $V$ is the Doppler centroid velocity in km\,s$^{-1}$ (values are not corrected for Earth's motion). Subscripts ``b'' and ``n'' indicate broad and narrow components, respectively. Quoted uncertainties correspond to $1\sigma$ statistical errors but systematic errors are not formally accounted here.}

\begin{tabular}{lcccccc}
\hline
 & \shortstack{$I_b$}
 & \shortstack{$I_n$}
 & $W_b$ $[\mathrm{km\,s^{-1}}]$
 & $W_n$  $[\mathrm{km\,s^{-1}}]$
 & $V_b$ $[\mathrm{km\,s^{-1}}]$
 & $V_n$  $[\mathrm{km\,s^{-1}}]$ \\
\hline
\hline

\multicolumn{7}{c}{Region A [$\mathrm{SB(H\beta_{n})}=3.39\pm0.03\times10^{-16}~\mathrm{erg\,s^{-1}cm^{-2}arcsec^{-2}}$]}\\
\hline
H$\beta$  &   $60\pm3$   & 100 & $1260\pm40$  & $190\pm20$   & $137\pm19$     & $248\pm1$     \\
H$\alpha$ & $221\pm5$    & $335\pm4$ & $1360\pm10$  & $127\pm12$   & $179\pm11$     & $278\pm1$     \\
He\,\textsc{i}(7065\,\AA) & $4.9\pm0.6$  & $0.9\pm0.2$ & $1700\pm500$ & $160\pm30$ & $150\pm60$ & $331\pm13$ \\
He\,\textsc{i}(5876\,\AA) & $1.6\pm0.4$  & --  & $500\pm300$  & -- & $350\pm50$ & -- \\
He\,\textsc{i}(7281\,\AA) & --   & $1.7\pm0.2$ & -- & $191\pm12$ & -- & $340\pm50$ \\
He\,\textsc{ii}(8236\,\AA) & $0.7\pm0.2$ & -- & $340\pm140$ & -- & $430\pm30$ & -- \\
\hline
\multicolumn{7}{c}{Region B [$\mathrm{SB(H\beta_n)}=4.75\pm0.03 \times10^{-16}~\mathrm{erg\,s^{-1}cm^{-2}arcsec^{-1}}$]}\\
\hline
H$\beta$ & $61.6\pm1.8$ & 100 & $1450\pm110$ & $170\pm20$ & $264\pm14$ & $248\pm1$ \\
H$\alpha$ & $207\pm4$ & $328\pm2$ & $1500\pm200$ & $122\pm17$ & $300\pm10$ & $278\pm1$ \\
He\,\textsc{i}(7065\,\AA) & $8.1\pm0.7$ & $0.4\pm0.2$ & $1500\pm200$ & $120\pm30$ & $390\pm50$ & $310\pm20$ \\
He\,\textsc{i}(5876\,\AA) & $4.0\pm0.7$ & -- & $660\pm130$ & -- & $290\pm40$ & -- \\
He\,\textsc{i}(7281\,\AA) & -- & $0.8\pm0.3$ & -- & $190\pm50$ & -- & $290\pm30$ \\
He\,\textsc{ii}(8236\,\AA) & $3.1\pm0.6$ & $0.3\pm0.1$ & $700\pm300$ & $81\pm4$ & $160\pm60$ & $234\pm12$ \\
\hline
\multicolumn{7}{c}{Region C [$\mathrm{SB(H\beta_n)}=2.03\pm0.01\times10^{-16}~\mathrm{erg\,s^{-1}cm^{-2}arcsec^{-1}}$]}\\
\hline
H$\beta$ & $32.8\pm2.6$ & 100 & $2500\pm200$ & $170\pm20$ & $420\pm60$ & $239\pm1$ \\
H$\alpha$ & $131\pm2$ & $357\pm5$ & $2430\pm100$ & $120\pm10$ & $480\pm30$ & $269\pm1$ \\
He\,\textsc{i}(7065\,\AA) & $8.2\pm0.9$ & $0.4\pm0.2$ & $2050\pm170$ & $160\pm70$ & $350\pm70$ & $780\pm20$ \\
He\,\textsc{i}(5876\,\AA) & $1.2\pm0.6$ & -- & $600\pm300$ & -- & $340\pm170$ & -- \\
He\,\textsc{i}(7281\,\AA) & -- & $0.7\pm0.1$ & -- & $134\pm10$ & -- & $248\pm12$ \\
He\,\textsc{ii}(8236\,\AA) & $0.4\pm0.8$ & $1.4\pm0.3$ & $520\pm140$ & $140\pm30$ & $270\pm60$ & $280\pm20$ \\
\hline

\end{tabular}

\end{table*}

\subsection{N103B}
\begin{figure*}
    \centering
    \includegraphics[width=16cm, trim={380 20 180 20}, clip]{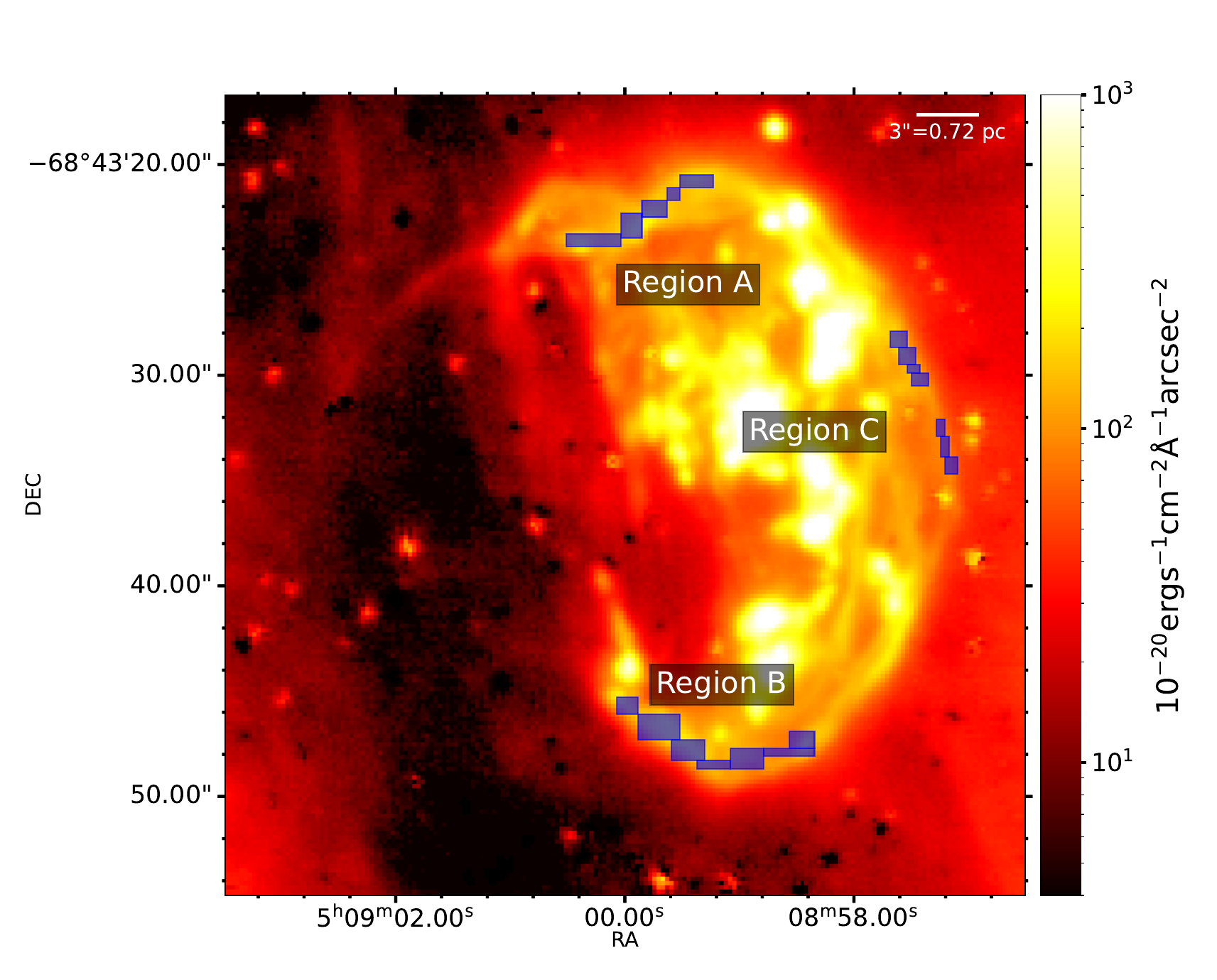}
    \caption{Continuum-subtracted mean H$\alpha$ flux density map (6473--6648\,\AA) of N103B from the MUSE datacube. A logarithmic stretch is applied for display with limits indicated on the colour bar. NaN pixels are shown in black. The blue rectangles represents the ROIs used for spectral integration and analysis.}
    \label{fig:N103B}
\end{figure*}

We performed a similar spectral analysis for three regions of N103B (Fig.~\ref{fig:N103B}; Table~\ref{tab:table2}). Both broad and narrow Balmer emission lines are evident, together with He~\textsc{i} features. Due to its location in the outskirts of the superbubble around the cluster NGC 1850, the spectra are contaminated by emission from the superbubble. Because of this, four- and five-Gaussian components were used to fit the H$\alpha$ emission line to accommodate the two nitrogen lines, as shown in Fig.~\ref{fig:nitrogen}. We also observe a slightly higher Balmer decrement (H$\alpha$/H$\beta$ > 4) and note that there is a small contribution to the Balmer emission lines from the superbubble. Correcting for the effects of LMC dust (de-reddening) is anticipated to cause a slight decrease in the Balmer decrement. Using an F-test, we identify an intermediate-width component in Regions A and B, in agreement with the intermediate H$\alpha$ line components reported by \cite{Ghavamian2017}, and we additionally find a new detection in Region C. We used the velocity width of the intermediate component of H$\alpha$ to constrain the bounds of our fit parameters for H$\beta$. Although the F-test does not formally require an intermediate component in H$\beta$, most likely because of the lower S/N in that line, we nevertheless report the constrained fit, since the clear detection of an intermediate component in H$\alpha$ makes its presence physically plausible in H$\beta$.

We report the detection of He \textsc{i} emission lines across all three regions in N103B, specifically He \textsc{i} 7065~\AA\ and He \textsc{i} 5876~\AA\ (see Fig.~\ref{fig:He_N103B}). Similar to SNR 0519, the helium lines exhibit much lower surface brightnesses compared to the hydrogen lines ($\sim$100 times lower than H$\beta$). The ratio of characteristic broad and narrow line components ($I_b/I_n$) in the hydrogen line profiles range from 0.1--0.3 which is somewhat less than reported by \cite{ghavamian17}.
The observed velocity width of hydrogen is significantly smaller (1000--1500 km\,s$^{-1}$) in Region B compared to Regions A and C ($\sim$2000 km\,s$^{-1}$). We detect and resolve both the broad and narrow component of He~\textsc{i} 7065~\AA\ across all three regions and broad He \textsc{i} 5876~\AA\ only in Region A. The presence of strong sky emission near He \textsc{i}\,5876~\AA\ in both Regions B and C inhibits extraction of the broad component. However, we detect narrow He~\textsc{i} 5876~\AA\ with measured velocity widths of $\sim$100--190 km\,s$^{-1}$.  Unlike SNR 0519, the width of the broad component of He \textsc{i} 7065~\AA\ is significantly smaller than that of hydrogen, ranging between $\sim$1200--1800 km\,s$^{-1}$.  We observe a similar dominance of the broad component over the narrow component in the helium lines, yielding \(I_b/I_n > 1\) in all regions of N103B.
The observed Doppler shifts for broad and narrow He~\textsc{i} and H are similar, with He having slightly higher Doppler velocities. A similar offset of $\sim$30--40 km\,s$^{-1}$ is again observed  between narrow H$\alpha$ and H$\beta$ across all regions. We do not, however, detect any He \textsc{ii} emission lines.

\begin{table*}
\caption{\label{tab:table2}Physical parameters of hydrogen and helium emission lines measured in N103B after correcting for Galactic reddening only (see Appendix~\ref{sec:appendixA}). The quantity $I$ gives the relative line intensity normalized to the narrow H$\beta$ component (by convention $I_{H\beta,n}=100$); the absolute surface brightness of narrow H$\beta$ is listed in the table header. $W$ denotes the velocity full width at half maximum (FWHM) of the fitted Gaussian components and $V$ is the Doppler centroid velocity in km\,s$^{-1}$ (values are not corrected for Earth's motion). Subscripts ``b'' and ``n'' indicate broad and narrow components, respectively. Quoted uncertainties correspond to $1\sigma$ statistical errors but systematic errors are not formally accounted here.}

\begin{tabular}{lcccccc}
\hline
 & $I_b$
 & $I_n$
 & $W_b$ $[\mathrm{km\,s^{-1}}]$
 & $W_n$ $[\mathrm{km\,s^{-1}}]$
 & $V_b$ $[\mathrm{km\,s^{-1}}]$
 & $V_n$ $[\mathrm{km\,s^{-1}}]$ \\
\hline
\multicolumn{7}{c}{Region A  [$\mathrm{SB(H\beta_n)}=8.2 \pm 0.2 \times10^{-16}~\mathrm{erg\,s^{-1}cm^{-2}arcsec^{-2}}$]}\\
\hline
H$\beta$ & $25.7\pm1.8$ & 100 & $1890\pm90$  & $168\pm10$  & $160\pm70$  & $220\pm1$ \\
H$\alpha$ & $85\pm5$ & $390\pm7$ & $2100\pm130$  & $120\pm8$   & $250\pm40$  & $251\pm1$ \\

He\,\textsc{i}(7065\,\AA) & $2.9\pm0.3$ & $0.22\pm0.08$ & $1750\pm490$ & $110\pm20$  & $270\pm90$  & $314\pm20$ \\
He\,\textsc{i}(5876\,\AA) & $3.8\pm0.1$ & $0.4\pm0.3$ & $1100\pm700$ & $100\pm80$ & $320\pm170$ & $310\pm30$ \\
\hline
\multicolumn{7}{c}{Region B  [$\mathrm{SB(H\beta_n)}=5.5\pm0.1\times10^{-16}~\mathrm{erg\,s^{-1}cm^{-2}arcsec^{-1}}$]}\\
\hline
H$\beta$ & $28\pm2.9$ & 100 & $1230\pm140$ & $164\pm12$  & $440\pm40$  & $226\pm2$ \\
H$\alpha$ & $126\pm4$ & $424\pm9$ & $1300\pm50$  & $120\pm10$  & $407\pm20$  & $255\pm1$ \\
He\,\textsc{i}(7065\,\AA) & $1.4\pm0.4$ & $0.6\pm0.2$ & $800\pm600$  & $126\pm12$  & $430\pm110$ & $318\pm13$ \\
He\,\textsc{i}(5876\,\AA) & -- & $2.2\pm0.2$  & --& $190\pm30$  & --& $313\pm11$ \\
\hline
\multicolumn{7}{c}{Region C  [$\mathrm{SB(H\beta_n)}=4.4\pm1.4 \times10^{-16}~\mathrm{erg\,s^{-1}cm^{-2}arcsec^{-1}}$]}\\
\hline
H$\beta$ & $19\pm5$ & 100 & $2170\pm160$  & $169\pm15$  & $340\pm40$ & $226\pm1$ \\
H$\alpha$ & $90\pm5$ & $377\pm6$ & $2250\pm140$ & $118\pm10$  & $350\pm50$ & $255\pm1$ \\
He\,\textsc{i}(7065\,\AA) & $2.1\pm0.6$ & $0.5\pm0.1$ & $1400\pm600$ & $130\pm20$  & $400\pm300$ & $325\pm20$ \\
He\,\textsc{i}(5876\,\AA) & -- & $2.5\pm0.2$ & --           & $148\pm15$  & --          & $312\pm6$ \\
\hline
\end{tabular}

\end{table*}

\subsection{SNR 0509}
\begin{figure*}
    \centering
    \includegraphics[width=16cm, trim={380 20 180 20}, clip]{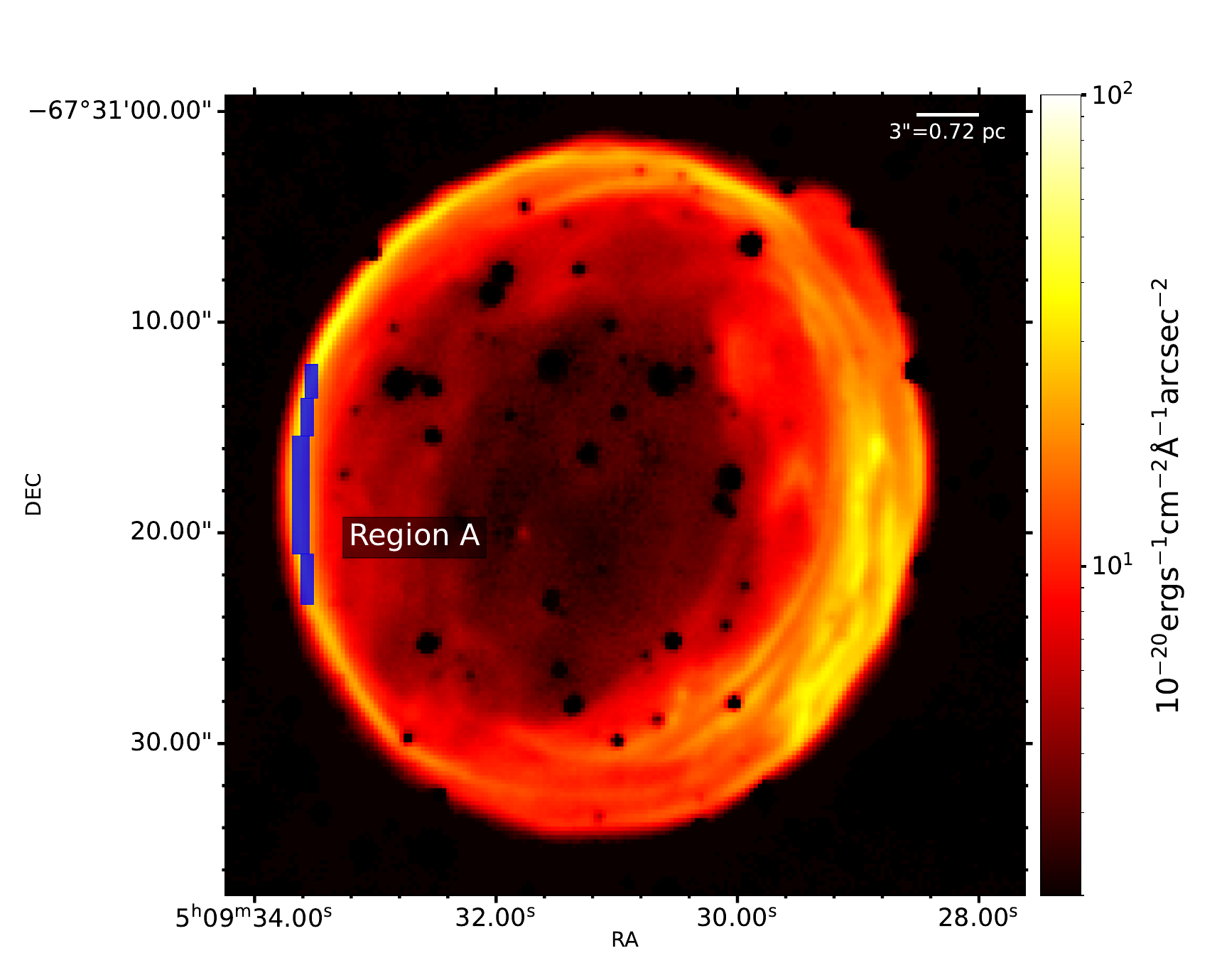}
    \caption{Continuum-subtracted mean H$\alpha$ flux density map (6464--6668\,\AA) of SNR 0509 from the MUSE datacube. A logarithmic stretch is applied for display with limits indicated on the colour bar. NaN pixels are shown in black. The blue rectangles represents the ROIs used for spectral integration and analysis.  }
    \label{fig:0509}
\end{figure*}

\begin{figure}
    \centering
        \includegraphics[width=0.5\textwidth, trim={0 0 0 0}, clip]{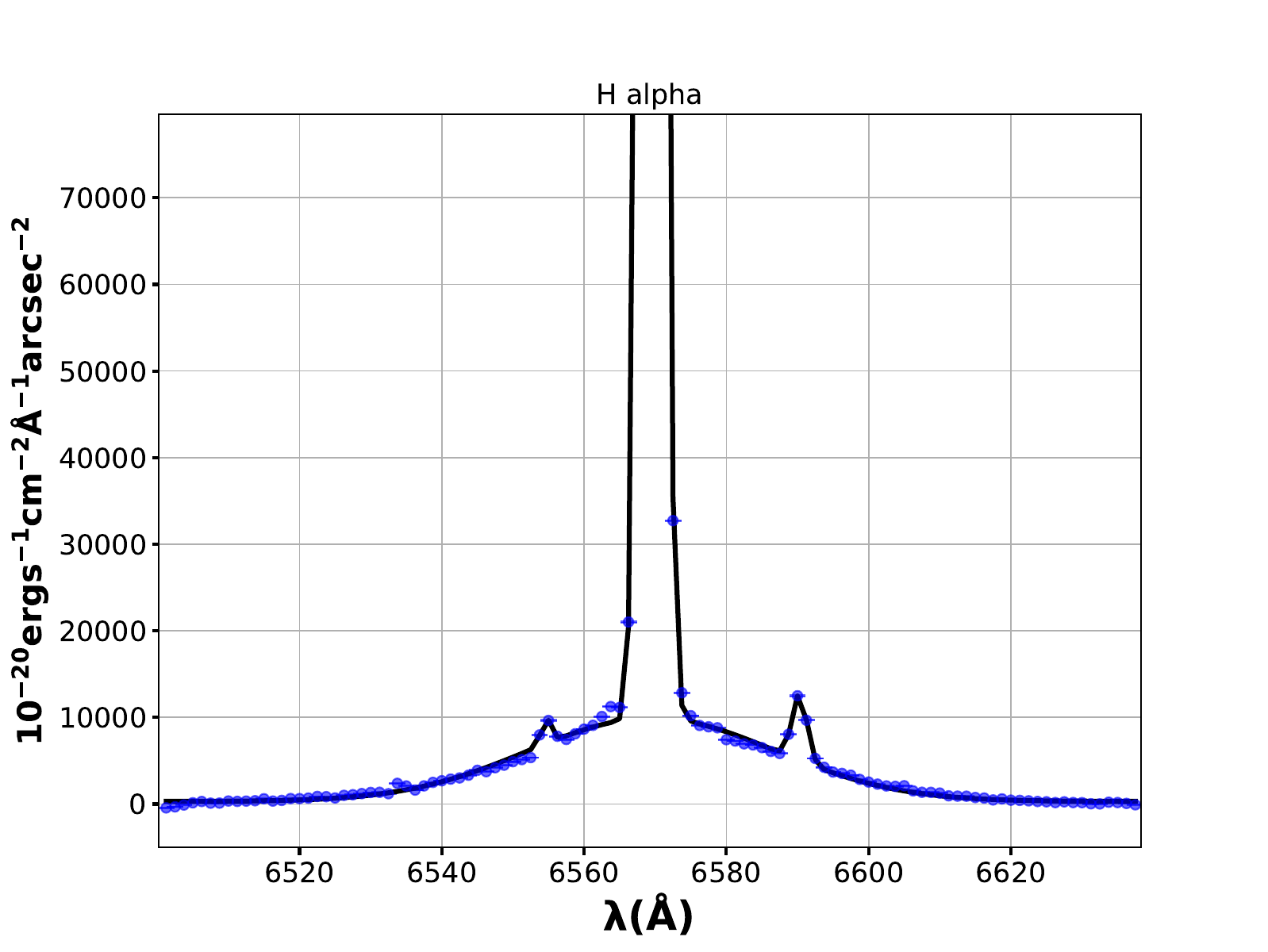}
        \includegraphics[width=0.5\textwidth, trim={0 0 0 0}, clip]{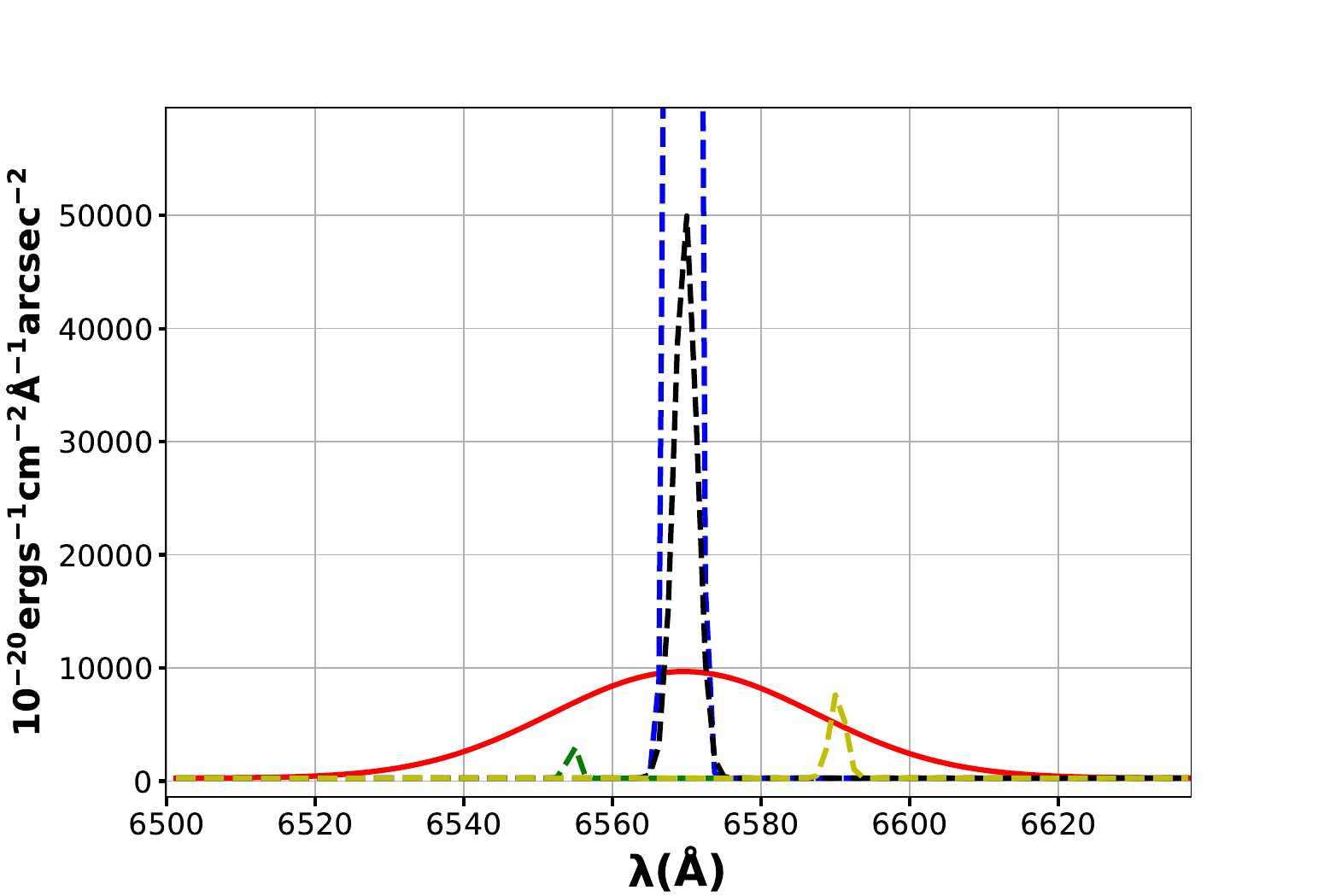}
    \caption{(a)  H$\alpha$ in Region A of N103B, shown by the blue data points with error bars, together with the total fit of five Gaussian components overlaid as a solid black line. (b) Visualization of the separate Gaussian fits. The green and yellow lines represent the fit to the nitrogen emission doublet.}
    \label{fig:nitrogen}
\end{figure}

\begin{figure*}
    \centering

        \includegraphics[width=0.5\textwidth, trim={0 10 40 60}, clip]{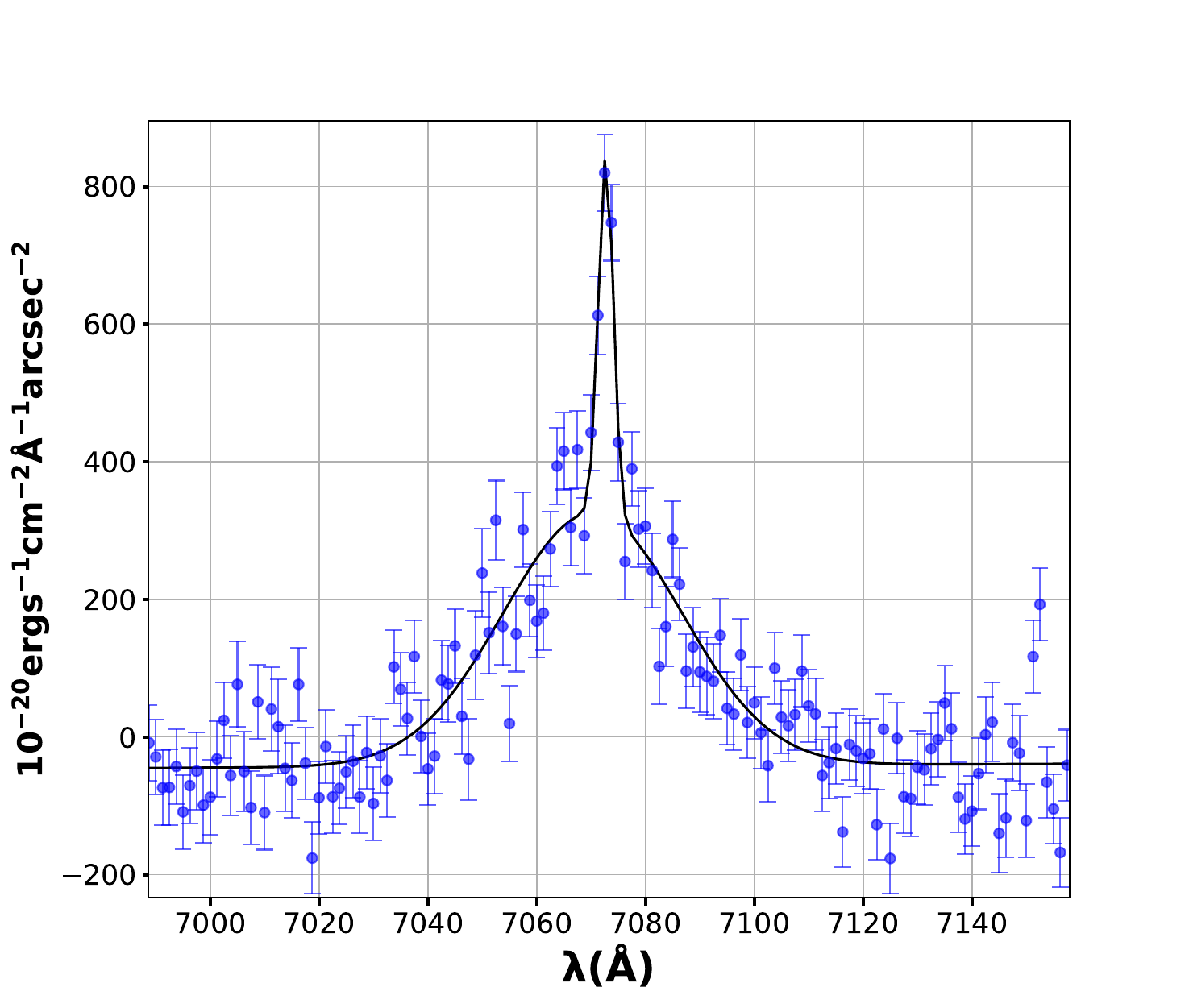}
        \includegraphics[width=0.5\textwidth, trim={0 10 40 60}, clip]{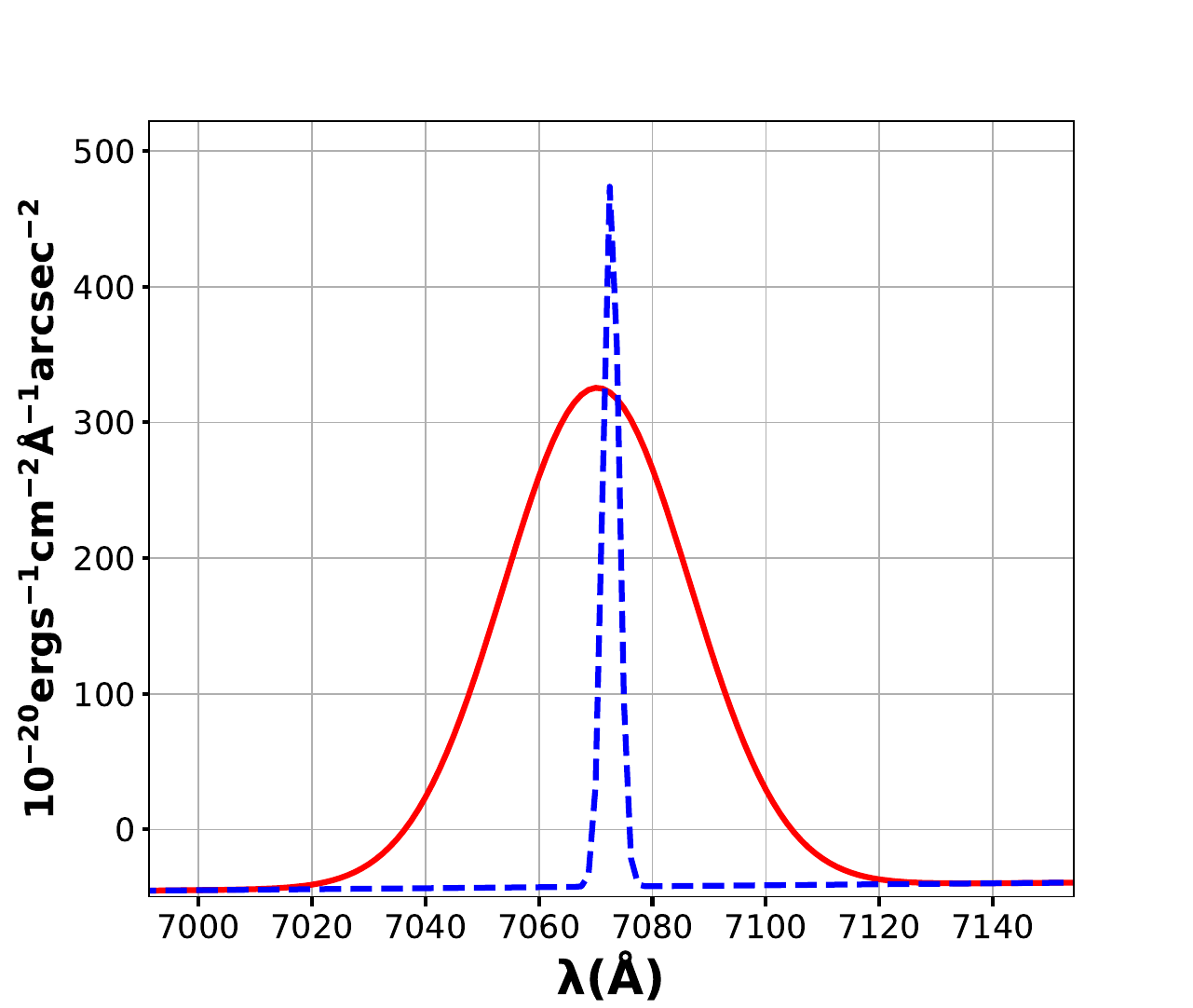}
    \caption{Broad and narrow Gaussian component fits for He~\textsc{i} 7065 \AA\ emission  from Region A of SNR 0519. Left: Data and combined fit. Right: Broad and narrow model components with the same linear continuum.}
    \label{fig:HeI_0519}
\end{figure*}

\begin{figure*}
    \centering

        \includegraphics[width=0.5\textwidth, trim={0 0 80 60}, clip]{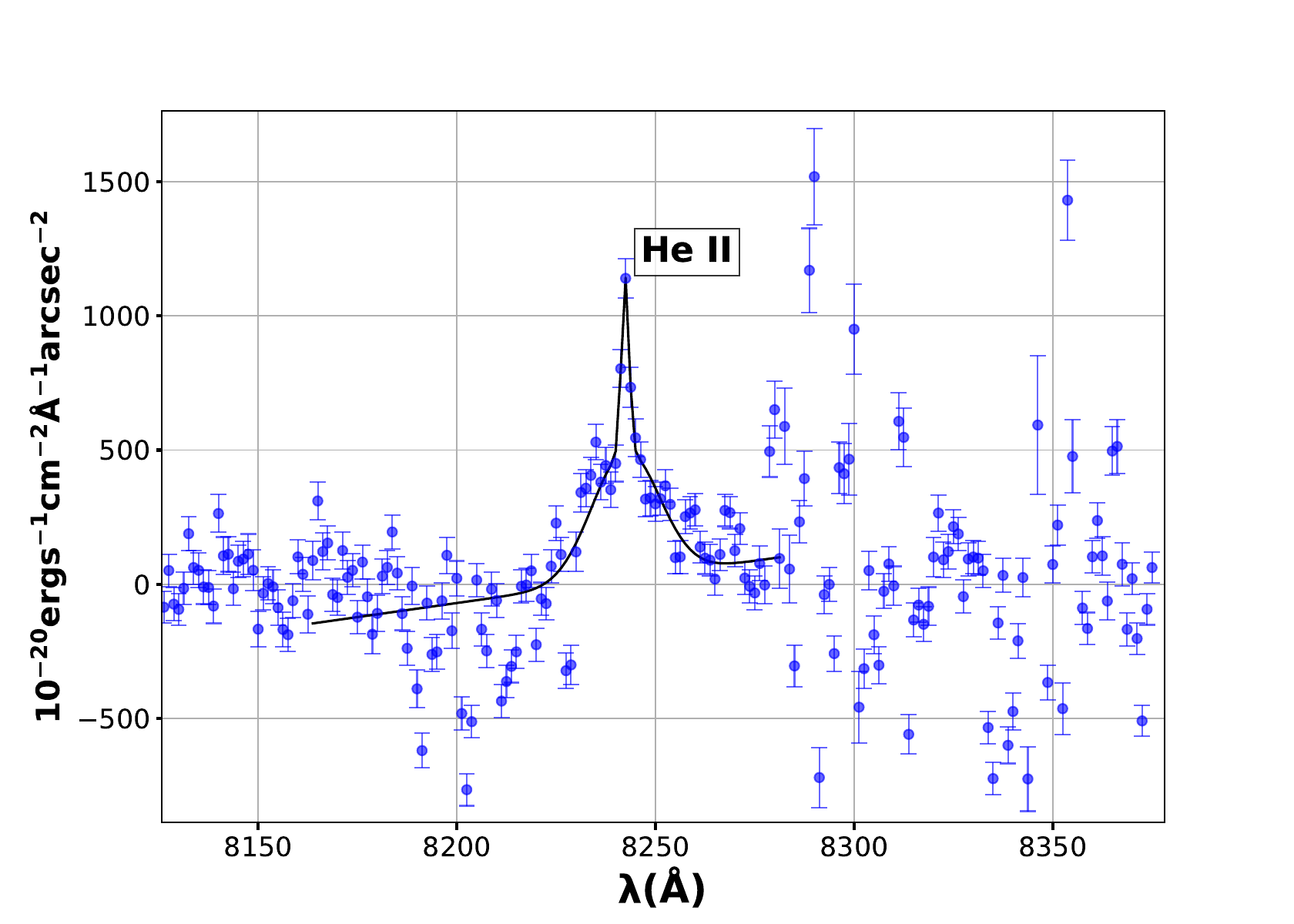}
    \hspace{0.005\textwidth} 
        \includegraphics[width=0.5\textwidth, trim={0 0 60 60}, clip]{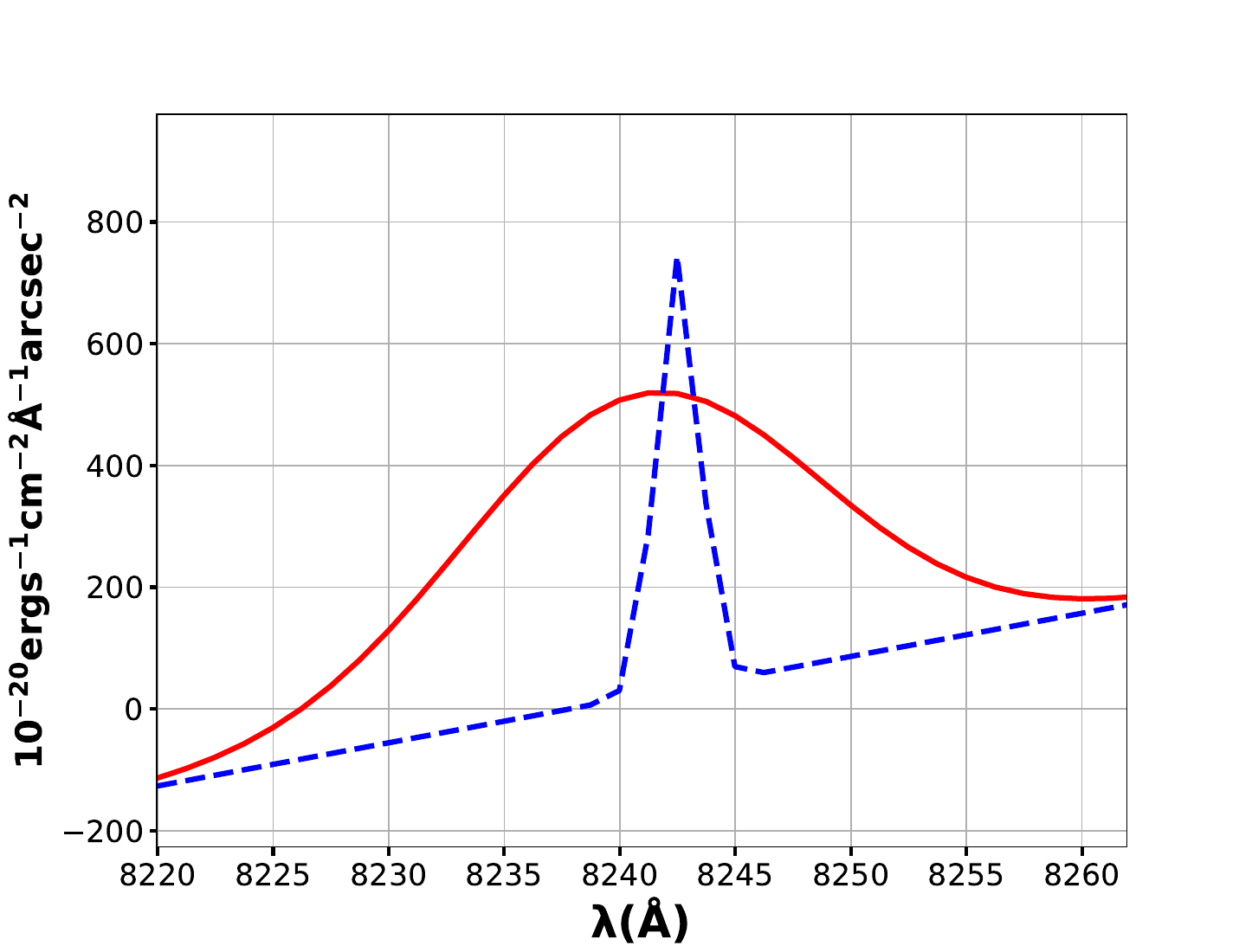}
    \caption{Broad and narrow Gaussian component fits for He~\textsc{ii} 8236 \AA\ emission from Region B of SNR 0519. Left: Data and combined fit. Right: Broad and narrow model components with the same linear continuum.}
    \label{fig:HeII_8236}
\end{figure*}

\begin{figure*}
    \centering

        \includegraphics[width=0.5\textwidth, trim={0 0 40 40}, clip]{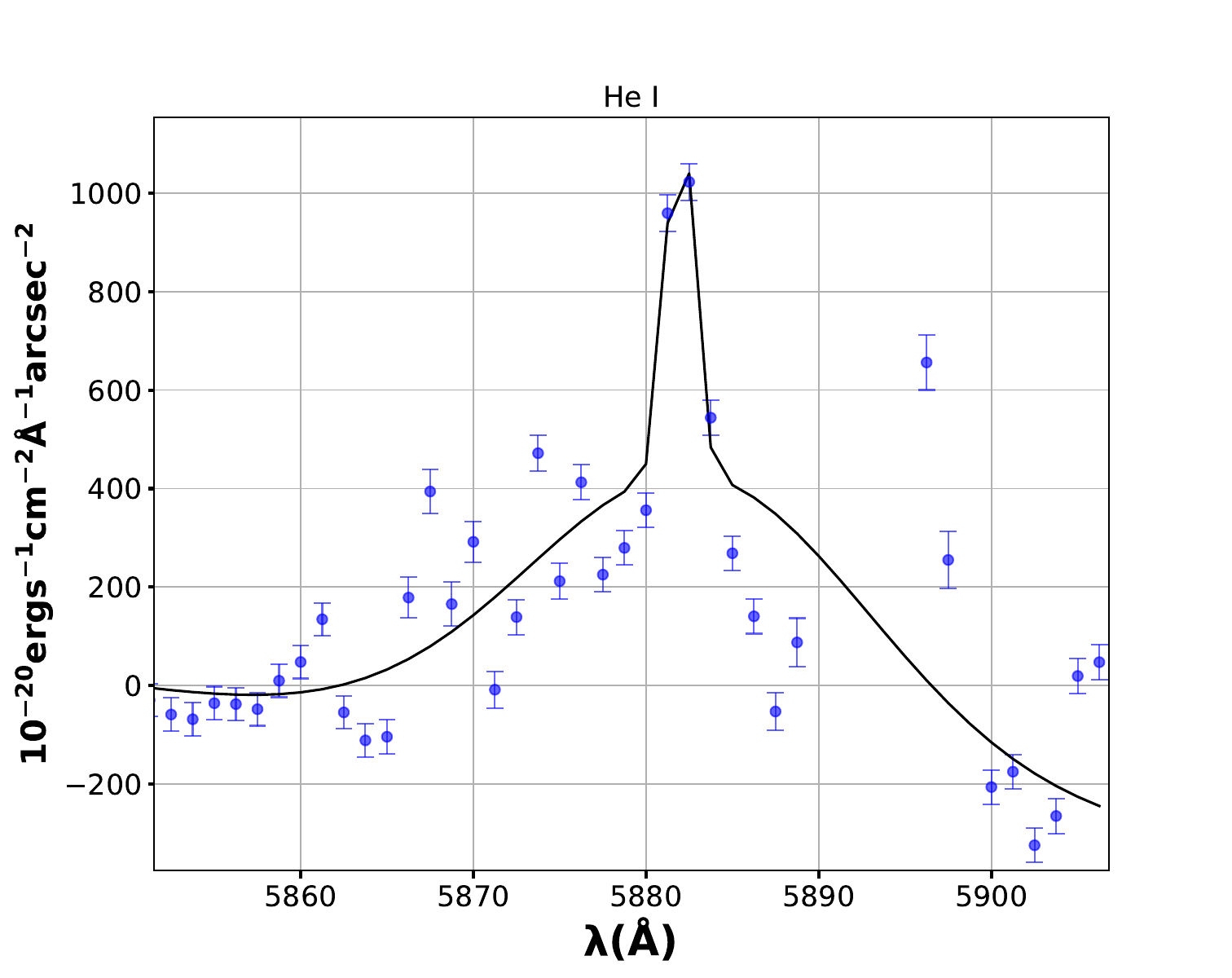}
    \hspace{0.005\textwidth} 

        \includegraphics[width=0.5\textwidth, trim={0 0 60 60}, clip]{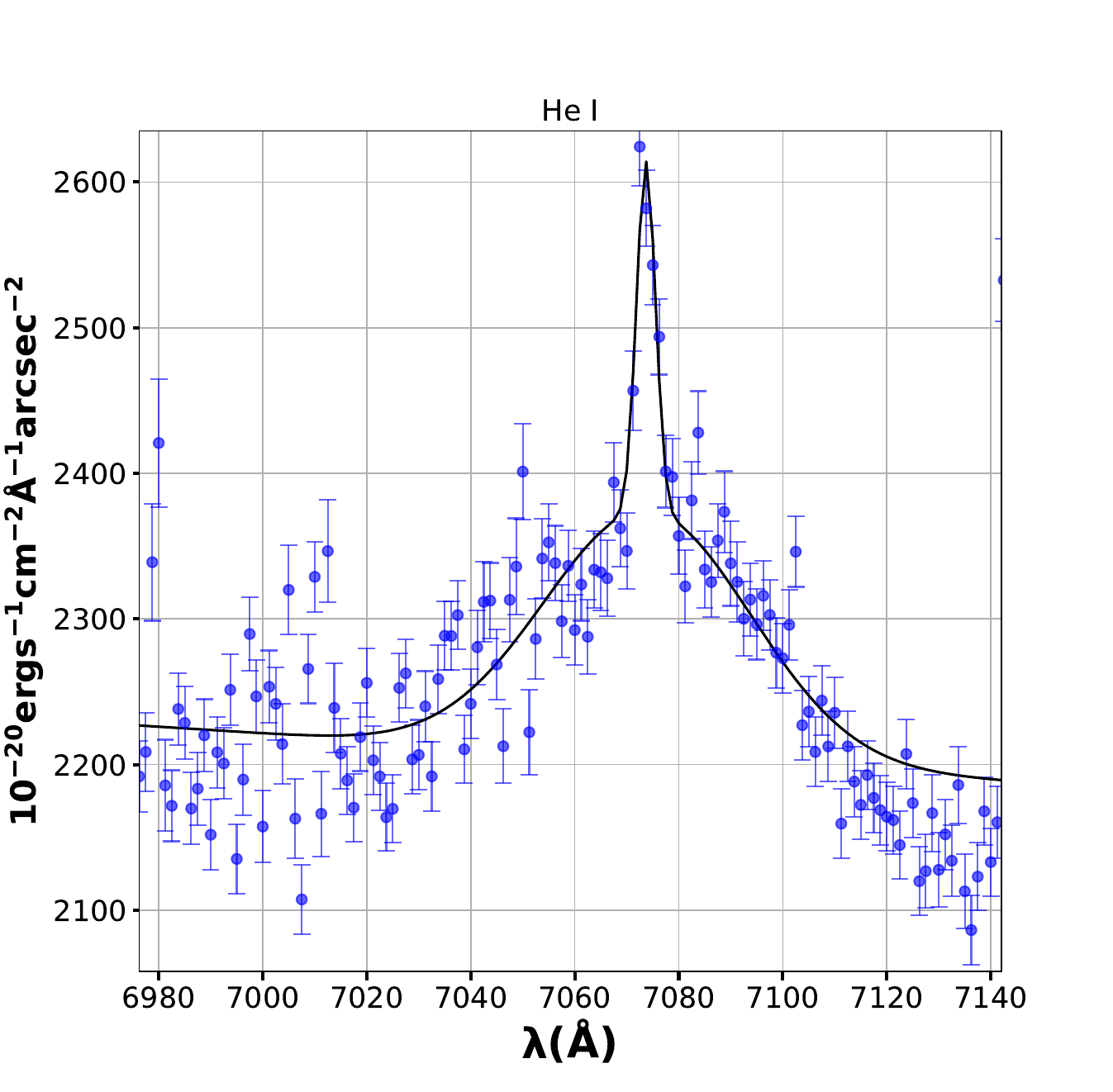}

    \caption{Broad and narrow Gaussian component fits for  He~\textsc{i} emission lines from  different regions of N103B. Left: He~\textsc{i} 5876 \AA\ detected in Region A with data points and the combined Gaussian fit. Right: He~\textsc{i} 7065 \AA\ detected in Region C with data points and the combined Gaussian fit.}
    \label{fig:He_N103B}
\end{figure*}

\begin{figure*}
    \centering

        \includegraphics[width=0.5\textwidth, trim={0 0 40 40}, clip]{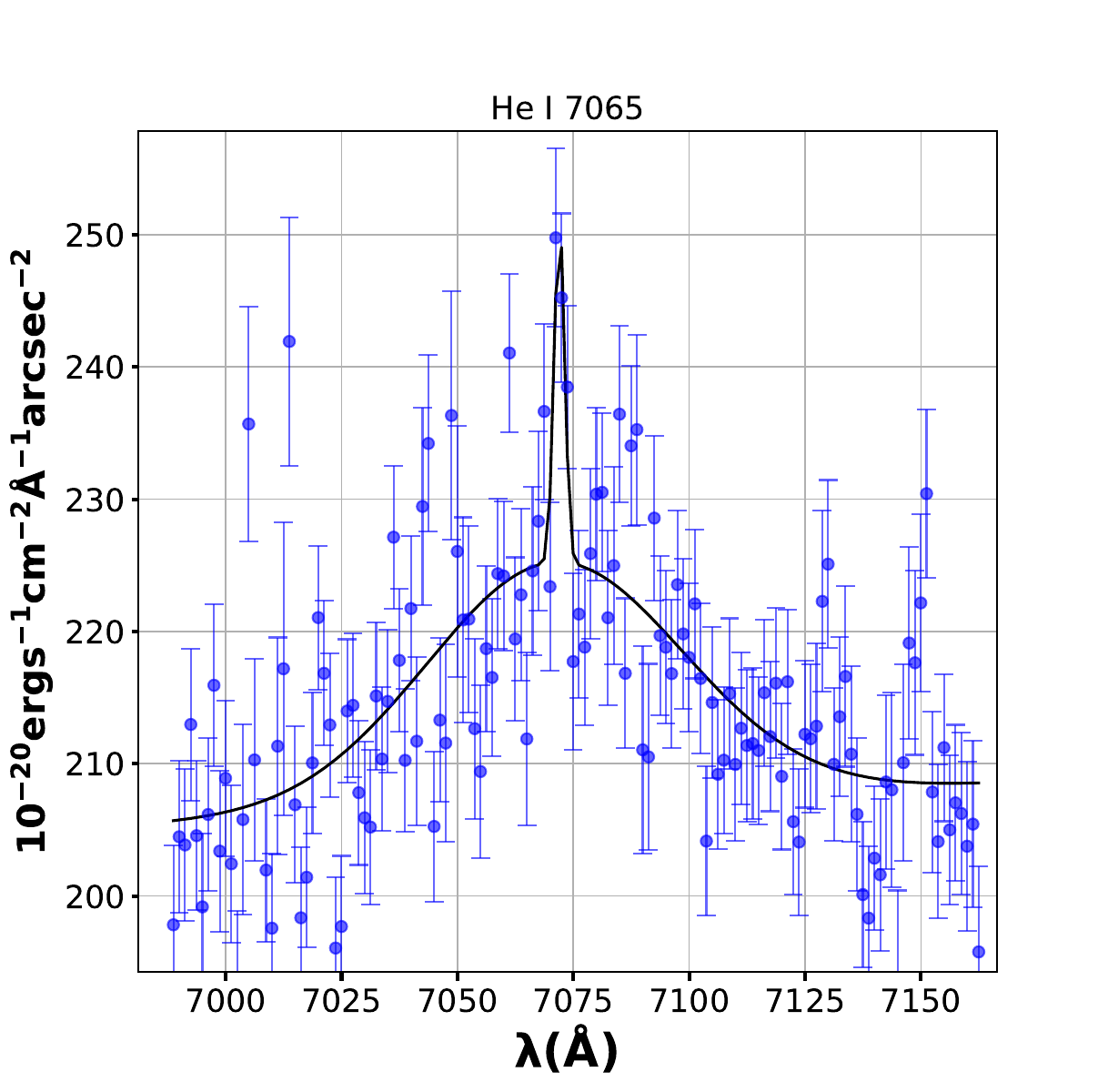}

    \hspace{0.005\textwidth} 

        \includegraphics[width=0.5\textwidth, trim={0 0 60 60}, clip]{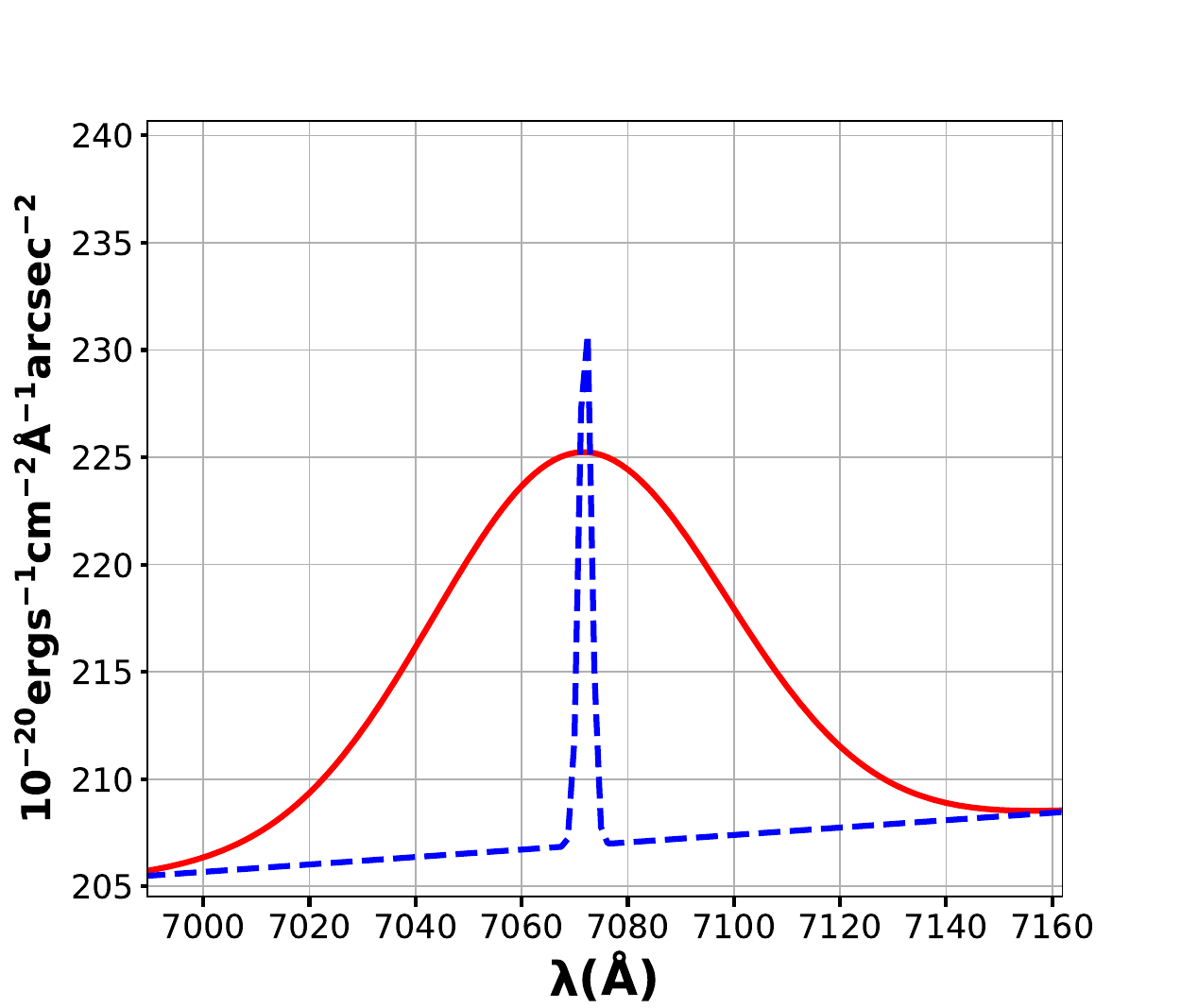}

    \caption{He~\textsc{i}~7065\AA\ emission line detected from  the eastern limb-brightened region of SNR~0509. Left: Data points and the combined fit. Right: Broad and narrow model components with the same linear continuum.}
    \label{fig:He_0509}
\end{figure*}

\begin{figure}
    \centering
    \includegraphics[width=9cm, trim={0 0 0 0}, clip]{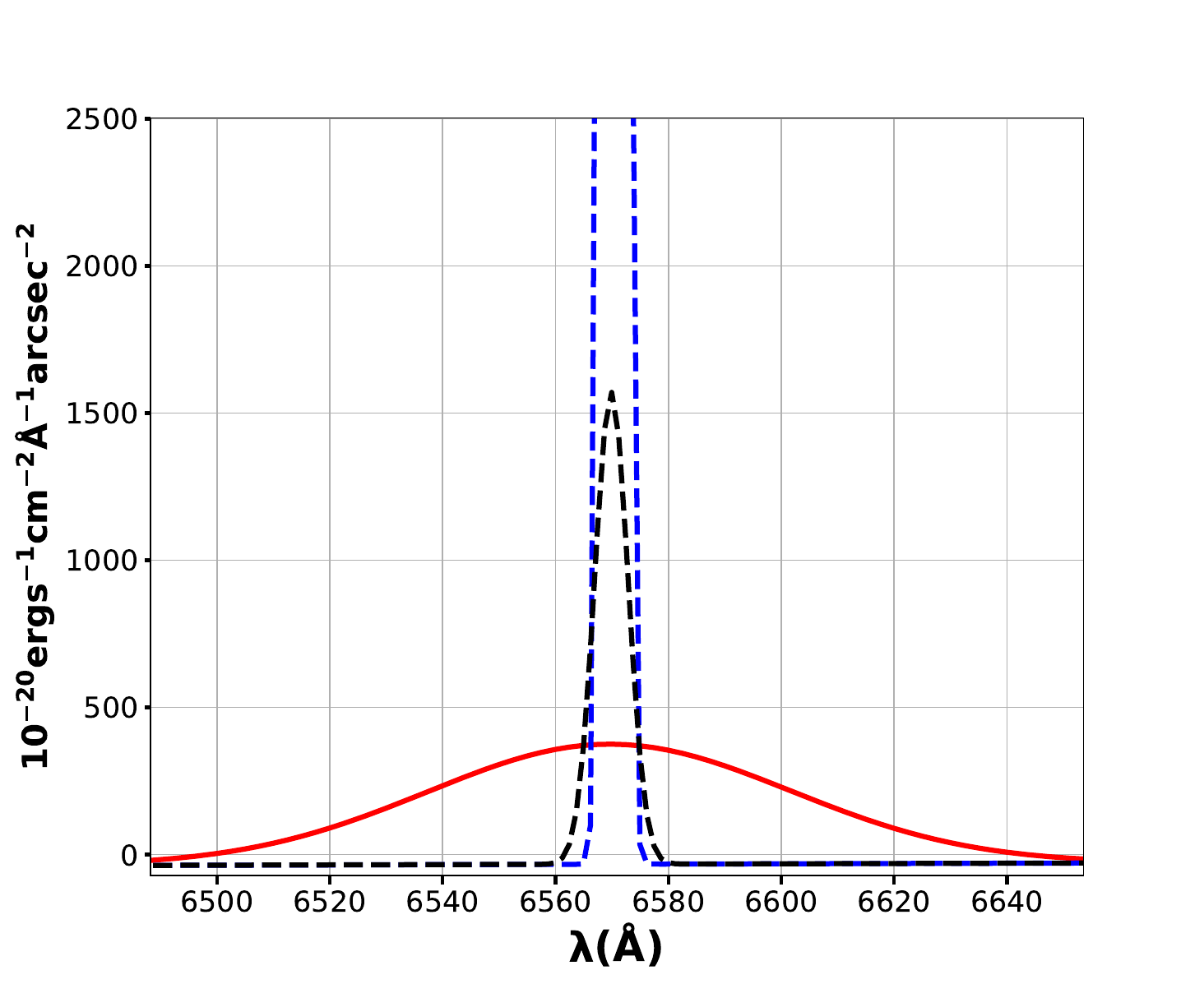}
    \caption{Broad, narrow and intermediate Gaussian component fit for the H$\alpha$ emission line in SNR 0509.}
    \label{fig:halpha_0509}
\end{figure}

The 30\,h deep observation of SNR 0509 reveals well-resolved H$\alpha$ and H$\beta$ lines, along with various faint He \textsc{i} emission lines. For our analysis, we selected a single region along the eastern limb-brightened edge of the remnant, as the western edge exhibits multiple superimposed filaments (see Fig.~\ref{fig:0509}). Because the western side shows this complex superposition of Balmer-dominated filaments and overlapping emission from coronal lines of the reverse-shocked ejecta, we refrain from analyzing the spectrum in that region.
 We report the detection of multiple He \textsc{i} lines alongside H$\alpha$ and H$\beta$ at 5015\,\AA, 6678\,\AA, 7065\,\AA\ and 7281\,\AA\, listed in Table~\ref{tab:table3} (also see Fig.~\ref{fig:He_0509}). An F-test indicated the presence of an intermediate component in H$\alpha$ and H$\beta$ (see Fig~\ref{fig:halpha_0509} and Table~\ref{table5}). The $I_b/I_n$ for the hydrogen lines is $\sim$0.07, similar to that reported by \cite{Hovey} and for He~\textsc{i}\,7065\,\AA\ $I_b/I_n > 1$, consistent with the other remnants in this paper. We detect both broad and narrow components for  He~\textsc{i}~7065~\AA\ with velocity widths of $\sim$2800 km\,s$^{-1}$ and $\sim$100 km\,s$^{-1}$, respectively. Only narrow components with widths of $\sim$100--150 km\,s$^{-1}$ are detected for the other helium lines, and we do not detect any  He~\textsc{ii}. The notch filter in the WFM AO mode excludes the wavelength range near He~\textsc{i}~5876~\AA. To investigate further, we examined an independent, earlier MUSE datacube of SNR~0509 (P.ID: 0100.D-0151[A], P.I.: Morlino) that did not employ the notch filter, but still found no detection of the line.
 The measured Doppler shifts based on our optimal parameters for the broad He~\textsc{i}~7065~\AA\ and H lines agree within the uncertainties, but the narrow He lines show a slightly greater Doppler shift compared to H. We again observe an offset of $\sim$30 km\,s$^{-1}$ between the Doppler shift of narrow H$\alpha$ and H$\beta$.

\begin{table*}
\caption{\label{tab:table3} Physical parameters of the hydrogen and helium emission lines measured in SNR 0509 after correcting for Galactic reddening only (see Appendix~\ref{sec:appendixA}). The quantity $I$ gives the relative line intensity normalized to the narrow H$\beta$ component (by convention $I_{H\beta,n}=100$); the absolute surface brightness of narrow H$\beta$ is listed in the table header. $W$ denotes the velocity full width at half maximum (FWHM) of the fitted Gaussian components and $V$ is the Doppler centroid velocity in km\,s$^{-1}$ (values are not corrected for Earth's motion). Subscripts ``b'' and ``n'' indicate broad and narrow components, respectively. Quoted uncertainties correspond to $1\sigma$ statistical errors but systematic errors are not formally accounted here.}

\begin{tabular}{lccccccc}
\hline
 & $I_b$
 & $I_n$
 & $W_b$ $[\mathrm{km\,s^{-1}}]$
 & $W_n$ $[\mathrm{km\,s^{-1}}]$
 & $V_b$ $[\mathrm{km\,s^{-1}}]$
 & $V_n$ $[\mathrm{km\,s^{-1}}]$ \\
\hline
\hline
\multicolumn{7}{c}{Region A [$\mathrm{SB(H\beta_n)}=1.84\pm0.04\times10^{-16}~\mathrm{erg\,s^{-1}cm^{-2}arcsec^{-1}}$]}\\
\hline
H$\beta$ & $7.5\pm1.1$ & 100 & $3660\pm470$ & $176\pm2$   & $220\pm80$ & $263\pm1$ \\
H$\alpha$ & $25\pm2$ & $350\pm8$ & $3310\pm200$ & $123\pm1$   & $280\pm60$ & $295\pm1$ \\
He\,\textsc{i}(5015 \AA) & -- & $1.17\pm0.05$ & -- & $145\pm7$  & -- & $400\pm3$ \\
He\,\textsc{i}(6678 \AA) & -- & $0.14\pm0.01$ & -- & $90\pm11$ & -- & $350\pm5$ \\
He\,\textsc{i}(7065 \AA) & $0.66\pm0.14$ & $0.05\pm0.02$ & $2260\pm430$ & $130\pm50$ & $280\pm140$ & $290\pm20$ \\
He\,\textsc{i}(7281 \AA) & -- & $0.59\pm0.06$ & -- & $100\pm30$  & -- & $360\pm6$ \\
\hline
\end{tabular}

\end{table*}

\section{He/H Abundance Ratios}
He/H abundance ratios measured at parsec scales around young Type~Ia SNRs provide a direct probe of the composition of the ambient medium, and therefore a powerful constraint on the progenitor system. These measurements directly connect the chemical environment a few parsecs from the explosion to the nature of the progenitor system itself, enabling discrimination between different evolutionary scenarios. Of particular interest are scenarios that require helium-rich companion (donor) stars, or more broadly, scenarios that involve companions that may have previously evolved through a helium-rich phase \citep[][see e.g. Fig 2]{liu2023a}. 

For reference, the standard primordial helium mass fraction is $Y_{\mathrm{P}}\simeq0.2454$ from Big Bang nucleosynthesis and CMB constraints \citep{Planck2020}. Converting this to a number ratio,
\begin{equation}
\frac{n_{\mathrm{He}}}{n_{\mathrm{H}}}
= \frac{Y_{\mathrm{P}}}{4(1-Y_{\mathrm{P}})}
= \frac{0.2454}{4(1-0.2454)}
\approx 0.0813,
\end{equation}
and so the primordial He/H abundance ratio by number is $\sim$8.1\%, providing a baseline for comparison with the helium abundances inferred from the BDS.

\subsection{He/H abundance ratio determination}
We use intensity ratios between the narrow components of H$\alpha$ and the various He\,\textsc{i} emission lines to estimate the He/H abundance ratio. In each case we write the line emission rate divided by the collisional ionization rate to give the number of photons emitted per atom going through the shock. For H\,\textsc{i} we follow the method in \citet{laming2001}. For He\,\textsc{i}, the collision rates are calculated by \citet{bray2000} and implemented in CHIANTI by \citet{delzanna2020}, from which we extracted the emission rates. Due to the different high energy dependencies of the various cross-sections, the number of photons per atom is temperature sensitive, and in the absence of multiple line ratios to allow a temperature diagnostic, an assumption about the electron temperature must be made.
The degree to which electrons and ions partition energy remains imperfectly known. Ghavamian et al. (2025, in preparation) determine $T_e/T_p < 0.1$ for many locations in SNR 0519 and SNR 0509, in agreement with the previous results of \citet{ghavamian2007b} and \citet{ Hovey}. They found a relationship, $T_e/T_p\sim 1/v_s^2$ for electron and proton temperatures $T_e$ and $T_p$, respectively, and shock velocity $v_s$. With $T_p\propto v_s^2$, this implies an approximately constant electron temperature with shock velocity, which evaluates to $\sim$$3\times 10^6$ K for SNR 0519.
We also assume $\sim$$3\times 10^6$ K here for N103B \citep{Someya2014,Heyden2002}, while SNR 0509 has sufficiently many line intensity ratios to allow the temperature to be determined independently.
In SNR 0519, the He\,\textsc{i}\,7065\,\AA\ to H$\alpha$ intensity ratio in Regions A, B, and C yields abundance ratios of 0.026--0.127, 0.029--0.143, and 0.017--0.084, respectively, taking collision rates at $3\times 10^6$ K. The ranges correspond to assuming Lyman $\beta$ is optically thin or thick. Assuming lower/higher temperatures would reduce/increase the inferred He/H abundance ratio. In N103B, the same analysis for Regions A, B, and C in that remnant yields ratios of 0.13--0.32, 0.12--0.29 and 0.33--0.81, again with the same sensitivity to  temperature.

A more conclusive He/H abundance ratio determination can be performed in SNR 0509. In Table~\ref{tab:table4} we give similar ranges for inferred He/H abundance ratios for a variety of temperatures. The He\,\textsc{i}\,5015~\AA\ /H$\alpha$ ratio is relatively insensitive to the assumed temperature, while the others exhibit some variation. This is to be expected, since the He\,\textsc{i}\,5015\,\AA\ line comes from an excitation to the $3 ^1$P term, for which the cross-section has a similar form to that for collisional ionization, while the other lines involving excitations to $3 ^1$D (6678\,\AA\,), $3 ^3$D (5876\,\AA\,), $3 ^1$S (7281\,\AA\,), and $3 ^3$S (7065\,\AA\,) have cross-sections of a different form. In fact, $3 ^1$D and $3 ^3$D are arguably also populated directly by charge exchange (see Section 6), making the He/H abundance ratios derived from these lines likely to be overestimates.
The best agreement between  He\,\textsc{i}\,5015\,\AA\ and He\,\textsc{i}\,7065\,\AA\ comes at a temperature of about $10^7$ K, where an He/H abundance ratio of 0.15--0.3 is indicated.
\begin{table*}
\caption{\label{tab:table4}  He/H abundance ratios across various electron temperatures for SNR 0509}
\begin{tabular}{ccccc}
\hline
& \multicolumn{4}{c}{$\log _{10} T_e$}\\
 Line & 5.75& 6.5& 7.0& 7.5\\ \hline
 
 He \textsc{i}(5015 \AA)& 0.146–0.346& 0.142–0.349& 0.193–0.297& 0.095–0.104\\
 He \textsc{i}(6678 \AA)& 0.010–0.023& 0.018–0.043& 0.035–0.054& 0.023–0.025\\
 He \textsc{i}(7065 \AA)& 0.006–0.013& 0.028–0.068& 0.104–0.160& 0.040–0.043\\
 He \textsc{i}(7281 \AA)& 0.047–0.110& 0.079–0.200& 0.150–0.230& 0.094–0.103\\ \hline

\end{tabular}
\end{table*}

\subsection{He/H abundance ratio discussion}
\label{sec:discuss-progenitor}

We emphasize that the He/H abundance ratios given above refer solely to neutral atoms. A true abundance ratio requires some consideration of the ionization balance. The cross-sections for photoionization are of similar order of magnitude \citep[e.g.][]{verner1996}, and photoionization of the surrounding medium by the progenitor or during the supernova event itself might leave similar fractions of neutral H and He. This seems to be borne out by calculations for SN 1006 by \citet{woods2018}, who also appear to rule out photoionization of the surrounding medium for SNR 0509. To understand and quantify the precursor effects on neutral H and He we used the photoionization precursor code described in \citet{Ghavamian2000b, Ghavamian2002} to constrain the preshock ionization fractions of H and He in  our objects of interest. We assume that most of the He ahead of the shock is initially neutral and consider two scenarios where the H\,\textsc{i} fraction is 0.5 \citep[see][realistic for SNR 0509]{ghavamian2007a,morlino13} and 0.9 (most generous case). The code models the photoionization of preshock gas by He\,\textsc{ii} 304\,\AA\, photons from the Balmer-dominated shocks.    Other sources such as He\,\textsc{ii} 2-photon emission and He\,\textsc{i} 584\,\AA\, emission (which can ionize both H~I and He~I) and He\,\textsc{i} 2-photon emission (which can ionize only H~I) contribute less than 5\% to the overall ionizing flux, and have negligible effect on the preshock ionization balance of hydrogen and helium. The code computes the neutral H and He fractions as a function of time and spatial position ahead of the shock. The model predicts that the He \textsc{i} fraction ahead of the shock is significantly reduced whereas the H\,\textsc{i} fraction remains fairly constant (see Table~\ref{tab:precursor_models}). This is due to the fact that He\,\textsc{ii} 304 \AA\ photons have energies near 41 eV, much closer to the ionization edge of He\,\textsc{i} (24 eV) than for H\,\textsc{i} (13.6 eV). So the cross-section is bigger for helium, and the energy deposited per photoionization is large. We use $f_{H\,\textsc{i}}$ and $f_{He\,\textsc{i}}$ values taking into account precursor ionization from Table~\ref{tab:precursor_models} (realistic assumption case, $f_{H\,\textsc{i}}=0.5$), assuming a He\,\textsc{ii} photoionization flux of a 2000 km/s shock in gas of density 1 cm$^{-3}$ (10$^7$ photons/cm$^2$/s, \citet{Ghavamian2000b}).   We evaluate the preshock neutral fractions at similar time frame of SNR 0509 ($\sim$300 years), SNR 0519 ($\sim$600 years) and N103B ($\sim$900 years) to constrain the effect of time variable neutral fractions on the He/H abundance ratio.\\ Let the total number of He be $Y$ and H be $X$, with $f_{He\,\textsc{i}}/f_{H\,\textsc{i}} =0.15-0.3$ from the observation for SNR 0509. This implies that $0.60Y/0.48X=0.15-0.3$ for SNR 0509 at 300 years, which indicates the total mass fraction of the He is 1.3--2.0$\times Y_P$ and He/H abundance ratio is 1.5--3.0 times the primordial He/H abundance ratio. Similarly for SNR 0519, the total He/H abundance ratio is 0.3--2.0 times the primordial value and for N103B the He/H abundance ratio is 1.7--3.4 times the primordial value. The above analysis indicates elevated He/H abundance ratios for SNR 0509 and N103B, even considering the full range of uncertainty. However, SNR 0519 shows an elevated He/h abundance ratio only when the upper limit is taken into account.

A particular scenario that could simultaneously give rise to a double-detonation explosion and at the same time leave an elevated He/H abundance ratio a few pc from the explosion site would be the so-called `ultraprompt' merger scenario of two white dwarfs predicted by binary population synthesis models \citep[see][section 5.2.1 and their figure~6]{ruiter2013}. This ultraprompt formation channel leads to mergers of double carbon-oxygen-rich white dwarfs that have delay times of $\sim$50--100 Myr and primary WD masses of ${\sim}$1  M$_{\odot}$ (with secondaries closer to ${\sim}$0.7 M$_{\odot}$).\footnote{We note here that in this ultraprompt merger channel, the initially less massive star on the ZAMS becomes the more massive, thus the primary, white dwarf star.} In such a scenario, the same star -- the initially less massive star on the ZAMS -- loses its envelope twice due to a common envelope (CE) event over the course of the binary's evolution; the binary additionally encounters one or more stable mass transfer events during evolution (see below). The second (final) CE event occurs when the initial primary is a WD and the initial secondary is a slightly evolved low-mass helium-burning star \citep[type 8 in the notation of][]{hurley2000} having already lost its H-rich envelope, and so this last CE is rich in helium, with an associated envelope mass of $\sim$1\,M$_{\odot}$.

\begin{table}[t]
\centering
\caption{Summary of 1D photoionization precusor models showing how He\,\textsc{ii} 304\,\AA\ ionizes H and He ahead of the shock at different times. We present two scenario with different initial neutral H ($f_{H\,\textsc{i}}$) and He ($f_{He\,\textsc{i}}$) fraction. $T_{\mathrm e}$ is the precursor electron temperature just upstream of the shock.}
\label{tab:precursor_models}
\begin{tabular}{ccccc}
\hline\hline
Initial $(f_{\mathrm{H\,I}}, f_{\mathrm{He\,I}})$ 
& Age (yr) 
& $f_{\mathrm{H\,I}}$ 
& $f_{\mathrm{He\,I}}$ 
& $T_{\mathrm e}$ (K) \\
\hline
(0.9, 0.9) & 300 & 0.86 & 0.60 & 18\,700 \\
(0.9, 0.9) & 600 & 0.80 & 0.40 & 25\,200 \\
(0.9, 0.9) & 900 & 0.70 & 0.25 & 26\,200 \\
\hline
(0.5, 0.9) & 300 & 0.48 & 0.60 & 14\,200 \\
(0.5, 0.9) & 600 & 0.46 & 0.40 & 19\,000 \\
(0.5, 0.9) & 900 & 0.43 & 0.30 & 21\,500 \\
\hline
\end{tabular}
\end{table}

These ultraprompt double white dwarf mergers are predicted to occur in {\tt StarTrack} models \citep{belczynski2008,ruiter2013} from relatively massive progenitors as far as SNe~Ia progenitors are concerned, with both stars having masses of $\sim$6\,M$_{\odot}$ on the ZAMS. The time delay between the final (He-rich) CE event and the WD merger resulting in the SN~Ia depends on the physical properties of the individual system, but is found to occur on the order of $\sim$$ 10^{5}$ ~yr in the models.

We can estimate approximately how far from the SNR we may expect to detect the He-rich material if it had been placed there as a consequence of the binary's final CE event. While CE speeds can be quite high, initially on the order of tens of km\,s$^{-1}$ \citep{roepke2023}, we expect the speed to decrease over time and thus we adopt an average CE velocity of 20 km\,s$^{-1}$ over the $10\,000$~yr `waiting period' plus the 300 yr remnant age. This gives a transverse distance of: 

\begin{equation}
    \left(\frac{20 \, \rm km}{\rm s} \right)  \left(10\,300 {\rm \, yr}\right)  \left( \frac{31\,557\,000 \, \rm s}{\rm yr} \right) \left( \frac{1 \,{\rm pc}}{3.086 \times 10^{13}\,  {\rm km}} \right) = 0.21 {\rm pc}
\end{equation}
While this distance is smaller than the $\sim$3.5\,pc to where we  observe the helium-enhanced material from the central region of SNR 0509, the order of magnitude agreement is encouraging, especially considering the lack of any assumptions about how the CE gas may become accelerated over the course of envelope ejection.

Although we have assumed a time delay of 10\,000~yr between the final CE event and WD merger based on population synthesis calculations, we emphasize that there would be a distribution of such times. It is also worthwhile to note that in the simulations, some evolutionary pathways include an additional mass transfer phase that occurs in which helium-rich material is transferred {\em stably} from the initially more massive star to the companion after the first CE event, but before the final CE takes place. While this final stable mass transfer event, though longer lived than the CE, could precede the final CE event by ${\sim}$$10^{5}$--$10^{6}$ yr, one may still expect to find the circumstellar material that is left surrounding the SN~Ia explosion to be enhanced in helium as a consequence. Further detailed studies would be required to assess the validity of the proposed formation channels, as there are a number of aspects to consider that go beyond the scope of this work. 

\begin{figure*}
    \centering
    \includegraphics[width=16cm, trim={0 0 0 0}, clip]{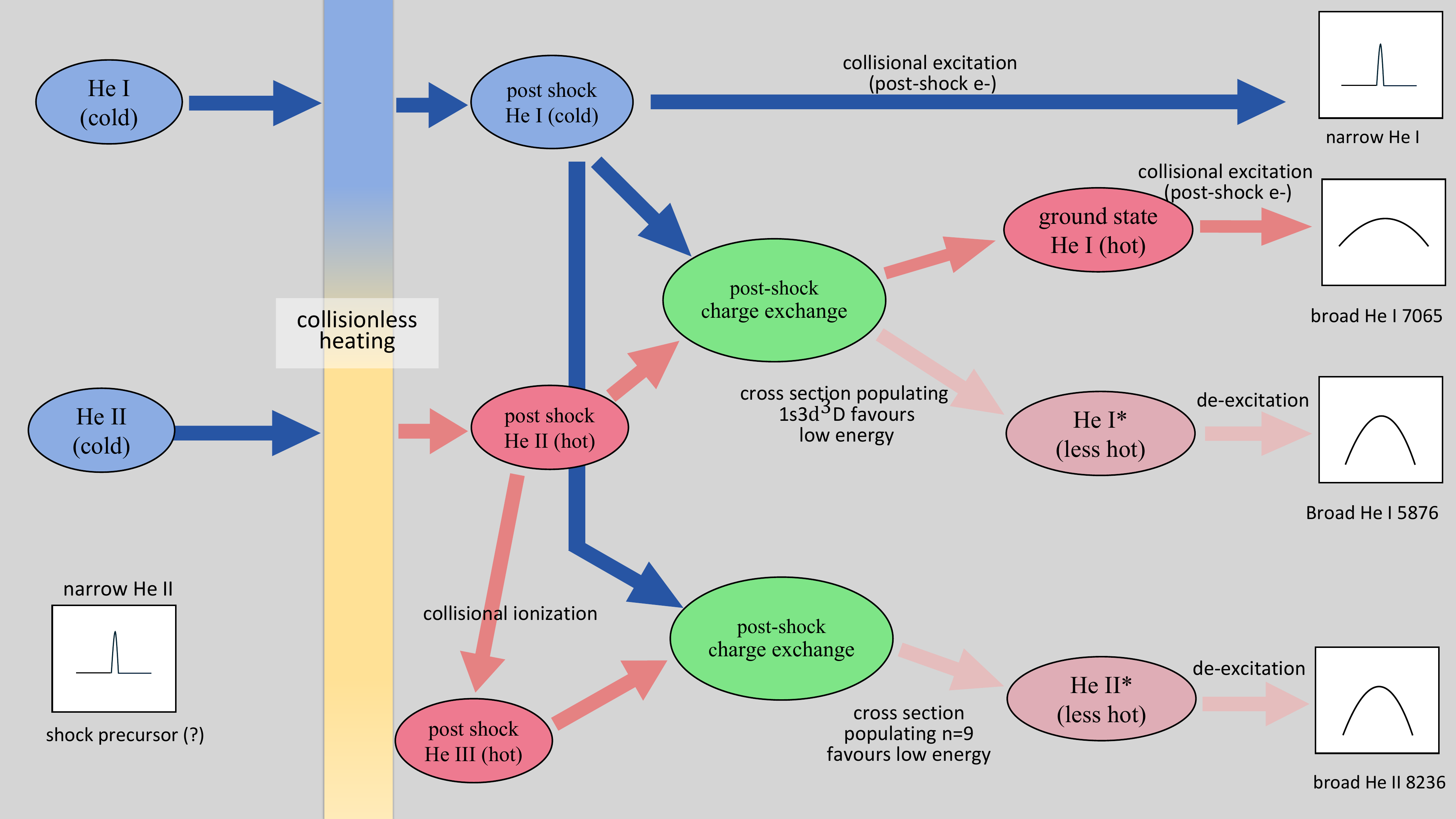}
    \caption{Schematic diagram explaining the formation of broad and narrow H~\textsc{i} and He~\textsc{ii} emission lines in the BDS region of Type Ia supernova remnants. Blue bubbles represents cold gas ahead of the shock and behind the shock produing narrow He~\textsc{i} emission lines. Green bubbles represents intermediate atomic processes. Red bubbles represent hot gasses producing broad He emission lines and pink bubbles represents less hot He~\textsc{i} and He~\textsc{ii} in the excited state producing similar less broad emission lines. Detection of Narrow He~\textsc{ii} is a puzzle and is assumed to be present ahead of the shock due to precursor ionization effects, however the exact mechanism is unknown (thus followed by a '?').}
    \label{fig:schematic2}
\end{figure*}

\section{Conclusions}
We report the first-ever detection of broad and narrow helium emission in the Balmer-dominated shocks of Type Ia supernova remnants in the LMC. The presence of broad He~\textsc{i} emission in all three LMC supernova remnants provides evidence for charge exchange between ionized helium and neutral helium in the post-shock region, analogous to the mechanism observed for hydrogen. This detection provides strong evidence for the formation of fast neutral He through this process (see Fig.~\ref{fig:schematic2}). Recent atomic calculations show that the charge transfer cross-section for He$^+$-He is similar to that of H$^+$-H ($10^{-15}~\mathrm{cm^2}$) at the same temperature \citep{Wang2023}. Since the abundance of He is an order of magnitude smaller than H, the calculations indicate a reduced broad-to-narrow ratio (B/N) of He than H. However, our observations of larger He B/N relative to H highlights the importance of charge transfer processes and complex scenarios such as branching into excited states, suppression of narrow H due to Lyman-Balmer conversion, or the presence of highly ionized He in the preshock region.

Although charge transfer represents a minor contribution and charge exchange between He~\textsc{i} and H$\alpha$ cannot be entirely ruled out, such a scenario is energetically disfavored.
A particularly intriguing case is SNR 0519, where He \textsc{i}~5876~\AA~ is detected without any narrow component. The width of this line closely matches the width of the broad He \textsc{ii} component, and  both are narrower than the He \textsc{i}~7065~\AA~ or the H \textsc{i} lines. He \textsc{i}~7065~\AA~ originates from the 1s3s $^3$S term which is populated by
electron impact excitation from the He ground state, once the fast He atom has formed when He$^+$ (already accelerated in the shock)
charge exchanges with an incoming neutral. He \textsc{i}~5876~\AA\ comes from the 1s3d $^3$D term, which has a much lower
electron impact excitation cross-section at the relevant electron impact energies \citep[by about an order of magnitude, see][]{ralchenko2008}. This upper level has a much higher statistical weight than  1s3s $^3$S, and so is more likely excited
directly by charge exchange. We speculate that such a reaction favours the lower velocity He$^+$ ions, meaning that the resulting
He~\textsc{i}~5876~\AA~ has a smaller thermal width than the He~\textsc{i}~7065~\AA, where electron impacts excite all atoms equally .
Some qualitative support for this idea can be found in the Oppenheimer-Brinkman-Kramers (OBK) approximation \citep{oppenheimer1928,brinkman1930} as summarized by \citet{Cocke2023}.

The detection of broad He~\textsc{ii} emission is a natural expectation across the remnants, arising from collisional excitation of ionized helium (from the pre-shock environment) behind the shock front. The 8236~\AA\ line is the $n=9$ to $n=5$ decay, and the $n=9$ state
is probably most efficiently populated by charge exchange or recombination. The fact that the line-width is very similar to that of 
He \textsc{i}~5876~\AA\ also suggests the same mechanism; He$^{2+}$ capturing an electron from an incoming neutral into an excited state and favouring lower velocity ions in the distribution function. The absence of He~\textsc{ii} in N103B and SNR 0509, as well as the non-detection of other transitional helium lines in SNR 0519, can likely be attributed to instrumental sensitivity limitations, especially given that the 8236~\AA ~line is the brightest and most readily detectable in these observations. 

The detection of a narrow He~\textsc{ii} component and its origin poses a significant challenge to our current understanding of atomic processes in collisionless shocks. We propose the hypothesis that the narrow He~\textsc{ii} arises from ionized helium formed by precursor radiation in the pre-shock region, indicating significantly ionized pre-shock gas.
In the post-shock region, ionized helium is expected to be thermalized through interactions with magnetic fields, reaching the same temperatures as other post-shock ions. This process should preclude the presence of narrow He~\textsc{ii} emission lines from the post-shock region. 
Moreover, our detection of He~\textsc{ii} 8236~\AA\ (9-5 transition) and undetected transition lines (10–5 at 7595 \AA, 11–5 at 7179 \AA, and 12–5 at 6899 \AA) highlights the complexity of the helium recombination cascade. Furthermore, the relative intensities of helium transitions vary markedly between the remnants: lines that are strongest in one object are comparatively weak in others. The observed line ratios often deviate from atomic physics predictions for transition strengths and simple recombination cascades, implying additional physical effects (e.g. non-equilibrium ionization, charge-exchange pathways, optical-depth or temperature-dependent population effects) are influencing the helium spectra.

Helium lines may present a new independent diagnostic tool for understanding the nature of the progenitor systems giving rise to Type Ia SNe. The He~\textsc{i} 7065\,\AA\ line emerges as the most reliable tracer of helium abundance across the three LMC remnants, showing consistency with He~\textsc{i} 5015\,\AA\ and 7281\,\AA. 
The discrepancy observed with He~\textsc{i} 5876\,\AA\ in N103B likely arises due to temperature-dependent ionization. The He\,\textsc{ii} 304\AA\, photoionization precursor from the Balmer-dominated shocks is expected to ionize neutral He more than H. However, the effect of ionization by ambient UV radiation and other potential ionizing effects were not taken into consideration. These effects might lower the neutral H fraction. SNR 0509 and N103B may indicate an enhanced He/H ratio under simple assumption, with a He/H abundance ratio (by number) of 12--24\% and 19--115\% respectively, 
particularly at higher temperatures ($\sim$$10^7$ K), suggesting temperature-dependent ionization balance. However, for SNR 0519 we get a He/H abundance ratio of 3--16\% indicating elevated helium only near the upper limit. Considering the effects, such enhancement is consistent with a helium-rich environment, which raises the prospect that the companion star for the progenitor of SNR 0509 might have been helium rich -- at least for some phase over the lifetime of binary evolution -- thus lending further support to a double-detonation scenario (see Section \ref{sec:discuss-progenitor}). Without temperature diagnostics or detection of more narrow He emission lines, precise abundance estimates remain uncertain for N103B, particularly given the temperature sensitivity of the He~\textsc{i} 7065\,\AA/H$\alpha$ ratio and the distinction between neutral and ionized helium populations. The enhanced He abundance found in our primary analysis is consistent with the results of \citet{Guest2022}, who concluded that N103B is interacting with dense circumstellar material. Nevertheless, pinpointing the specific process responsible for the increased helium is challenging, because N103B is located in the outskirts of the NGC 1850 cluster and observations are affected by diffuse cluster emission.

As discussed, the unique `ultraprompt' formation channel is at this point only a prediction from population synthesis models and remains to be confirmed by further studies both through simulations and observations. However, we propose that it is a plausible channel worth considering for linking several puzzle pieces that SNR 0509 has presented to the community \citep{Shields2023,Noda2016, Schaefer2012}. 
The specific binary system we identified in \texttt{StarTrack} calculations suggests an He-enrichment that extends to a distance of $\sim$0.2 pc, which is somewhat smaller than the $\sim$3.5 pc radius of the remnant. Although we have not identified a system that provides a complete quantitative reconstruction of the circumstellar environment that matches our new observational constraints, we note that this discrepancy may reflect the fact that we have examined only one representative system. A spread in binary parameters across similar systems could lead to a range of enrichment scales, and prior mass-loss episodes on $10^{5}$–$10^{6}$ year timescales (as discussed in Section 5) could also transport He to radii of order a few parsecs, consistent with the $\sim$3.5 pc scale of the forward shock. 

We have sketched how our new observational He/H ratio can be qualitatively connected to progenitor models and how such connections can yield real constraints on the progenitor systems and the nature of the companion. Establishing such links potentially provides a novel way to directly constrain  possible progenitor scenarios and mass-loss histories leading up to the explosion.
We introduced merging double white dwarf systems with ultrashort delay times as a physically motivated framework that could reconcile both the enhanced neutral He/H ratio in the ambient medium and mounting evidence that SNR 0509 originated from a sub-Chandrasekhar-mass progenitor via a double-detonation explosion \citep{Seitenzahll2019, Das2025, Mandal2025}. We therefore encourage more precise determinations of the ionization fraction in the preshock gas, detailed common-envelope simulations and deeper observations of Balmer-dominated filaments in supernova remnants to further test and refine this scenario.

\section{Acknowledgments}
JML was supported by Basic Research Funds of the Office of Naval Research, and thanks Dr. Ignacio Ugarte-Urra for assistance with Chianti computations.

\section*{Data Availability}
This work is based on archival observations obtained with MUSE on the VLT under P.IDs 096.D-0352(A) and 0104.D-0104(A). The raw data is publicly available from the ESO Science Archive \href{archive}{https://archive.eso.org/cms.html}. Additional reduced data and analysis scripts used in this study are available from the corresponding author upon request.

\bibliographystyle{aasjournal}
\bibliography{mainbib,snr}

\appendix

\section{Local background subtraction, De-reddening and Stellar continuum removal} \label{sec:appendixA}
\subsection{Local background subtraction}
\begin{figure}
    \centering
    \includegraphics[width=10cm, trim={380 20 150 20}, clip]{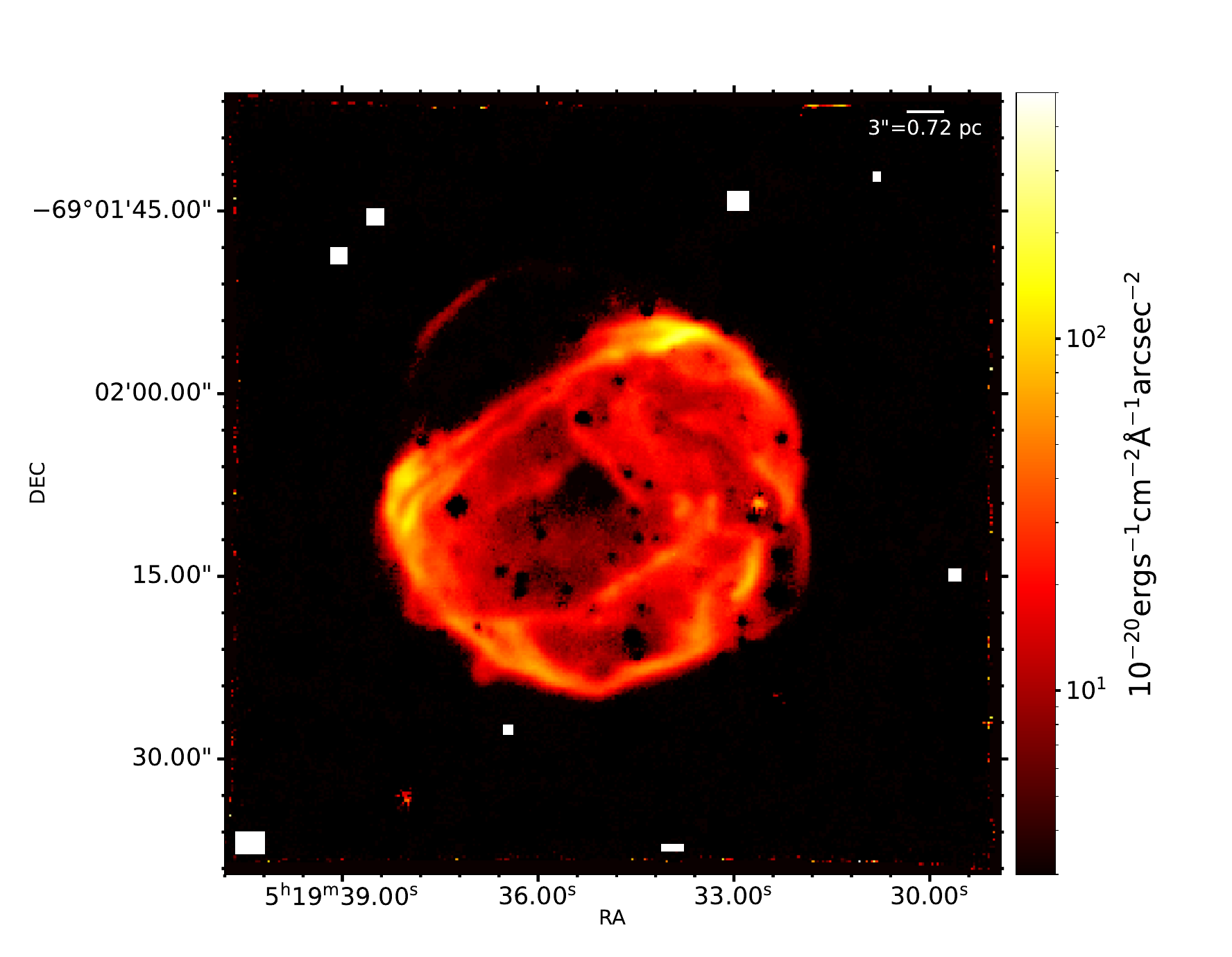}
    \caption{Sky background (white boxes) used for local background subtraction of SNR 0519.}
    \label{fig:bkg}
\end{figure}

\begin{figure*}
    \centering
    \includegraphics[width=20cm, trim={70 50 80 0}, clip]{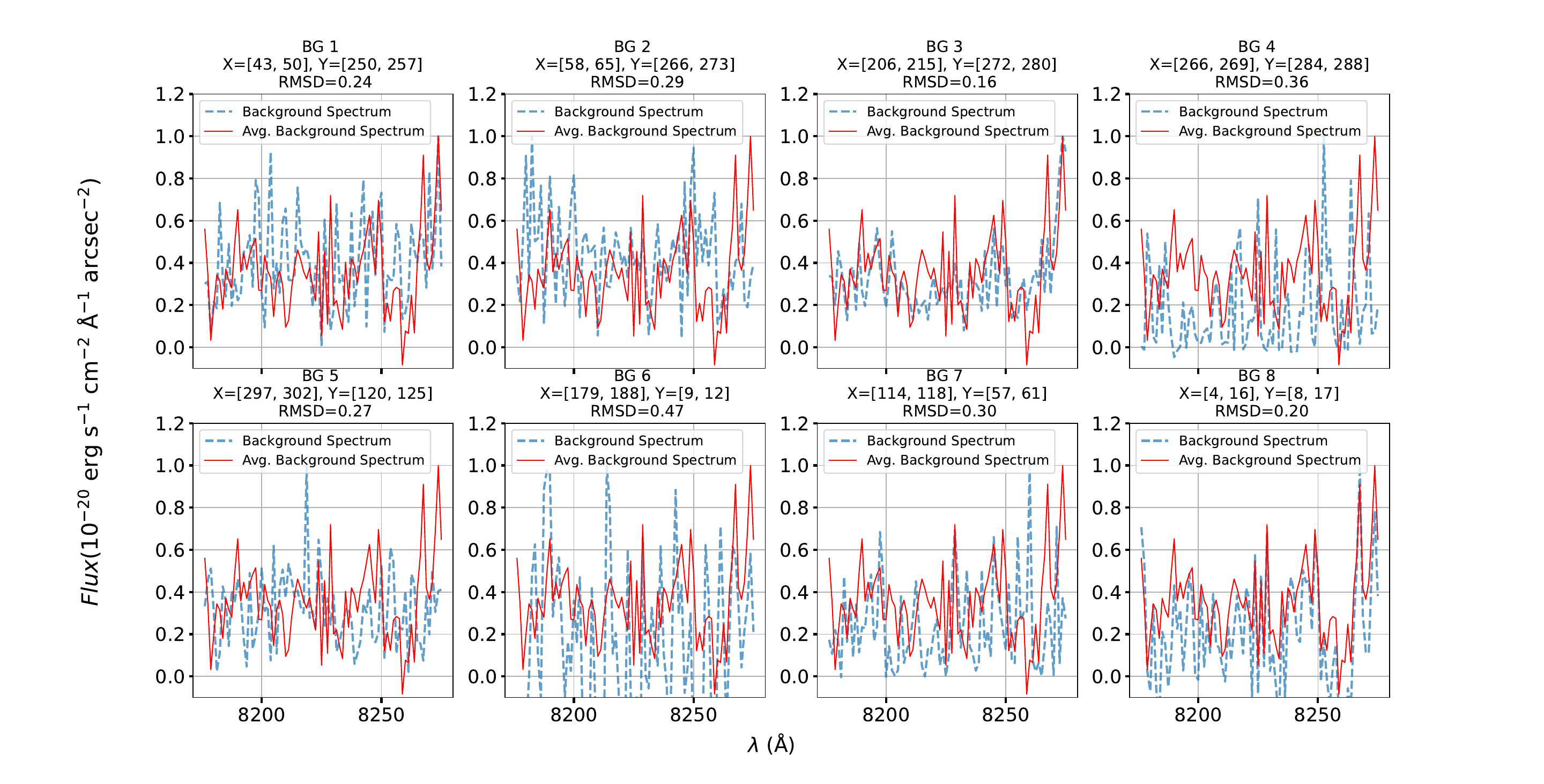}
    \caption{Spectra of individual background in the range  is plotted along with the spectra of the average background. RMS deviation values are plotted on the top of each spectra along with their spaxel coordinates}
    \label{fig:bkg_spectra}
\end{figure*}

To perform the local background subtraction, we selected eight small regions from the final data cubes  of SNR 0509 and SNR 0519 using the \texttt{QFitsView} tool (see Extended Data figure 2 of \citealt{Das2025} and Figure~\ref{fig:bkg}; \citealt{2012ascl.soft10019O}), such that these regions are away from any stellar emission. We create an average sky spectrum by combining emission dominated by skylines from all regions and then normalize the flux by dividing by the total number of pixels used to construct the spectrum. Fig~\ref{fig:bkg_spectra} plots the RMS value of deviation between spectra of individual background regions in SNR 0519 and the average background.
A cube of the same shape as the deep datacube is then subtracted from the original cube. We did not perform any such local background sky subtraction for N103B due to its location in the superbubble which produces strong H$\alpha$ and nitrogen emission lines with varying flux throughout the cube.  Implementing a local background subtraction leads to an over subtraction of the nitrogen lines, producing artificial absorption features. The hydrogen background emission is approximately 25--30 times fainter than the Balmer emission from the remnant and therefore does not significantly affect our analysis.

\subsection{Galactic extinction}
\begin{figure}
    \centering
    \includegraphics[width=9cm, trim={0 0 0 0}, clip]{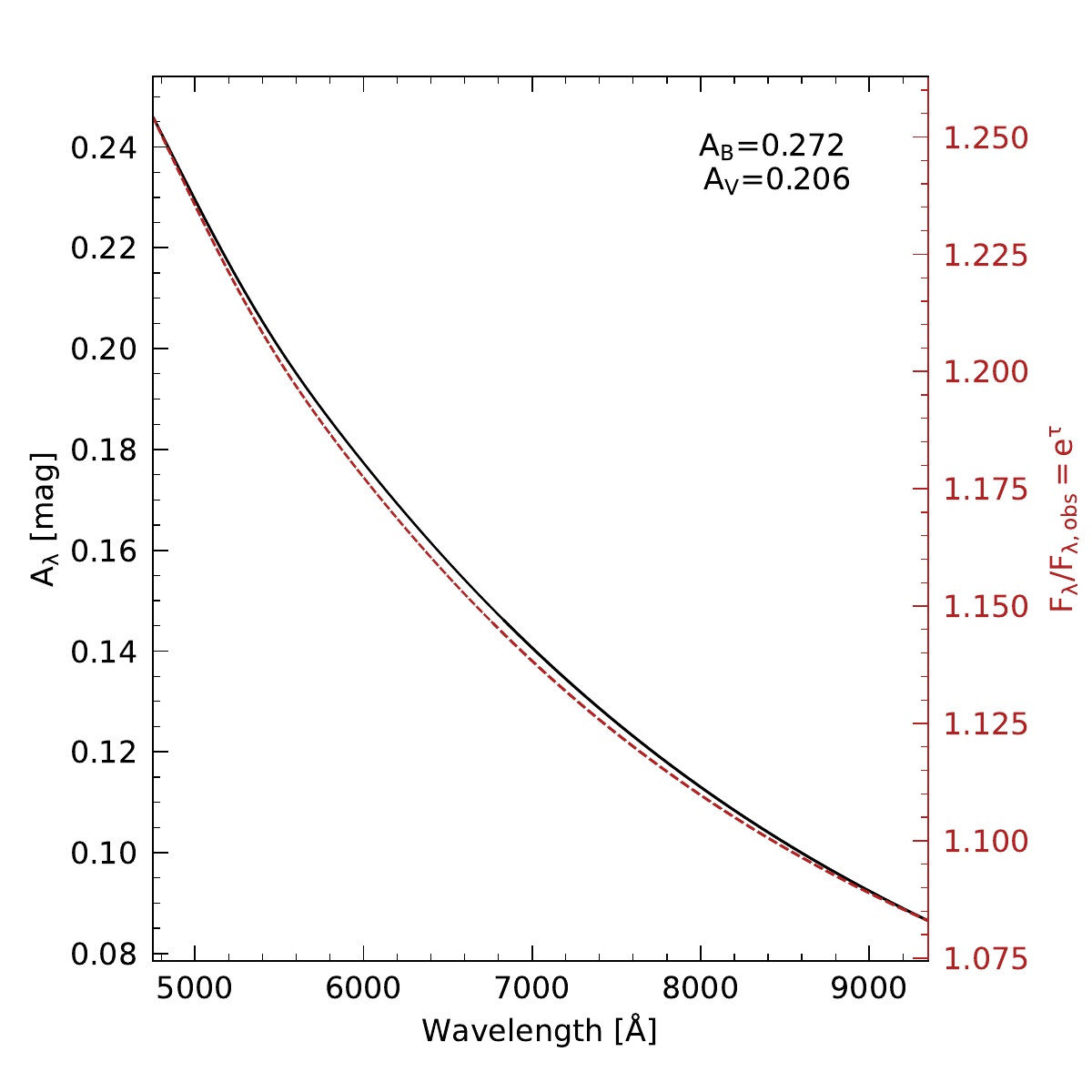}
    \caption{Extinction curve for optical wavelengths produced by \texttt{Brutifus}. This curve was used to produce de-reddened spectra for each cube. The black line shows the absolute extinction as a function of wavelength, while the red line shows the flux correction factor as a function of wavelength.}
    \label{fig:dered}
\end{figure}

Light coming from the targets to the detector is affected by interstellar dust. Dust will preferentially scatter the shorter wavelengths of light, leaving the longer (redder) wavelengths. It is therefore necessary to correct for reddening effects.  The   \cite{fitzpatrick1999correcting} reddening law with $R_V$ = 3.1, $A_B$ = 0.272, and $A_V$ = 0.206 was applied, with these values obtained from NED \citep{schlafly2011measuring} based on a re-calibration of the infrared-based dust map \citep{schlegel1998maps}. Since all the targets are located in the LMC, these Galactic values are sufficient for this work. The combined datacubes for all three targets have been corrected for Galactic extinction along the line-of-sight using a custom  \texttt{Brutifus} (https://github.com/brutifus) procedure and the above values, as shown in Fig~\ref{fig:dered}.  The output from the de-reddening recipe was used as an input for the continuum subtraction recipe in \texttt{Brutifus}. Stellar continuum was effectively subtracted using the tool by fitting a Lowess curve. We used 10 iterations with a Lowess fraction of 0.05 for sigma-clipping to eliminate outliers. We note that there will be residual LMC reddening affecting our observations. However, because the exact location of the remnants within the LMC is uncertain, it is difficult to compute the exact reddening contribution by the fraction of LMC dust in our line-of-sight.

\section{F-test, calculated parameters and error propagation}\label{sec:appendixB}
\subsection{F-test implementation}
We used both two and three Gaussian components to investigate the existence of an intermediate component for H$\alpha$ and H$\beta$ emission lines in the BDS regions of SNR 0509 and 0519. An F-test on the number of Gaussian components was performed to ensure that fitting an intermediate component substantially increased the quality of the fit.  The F-statistic is given by:
\begin{equation}
F = \frac{(RSS_1 - RSS_2) / (k_2 - k_1)}{RSS_2 / df_2}
\end{equation}
where:
\begin{itemize}
    \item $RSS_1$ and $RSS_2$ are the residual sum of squares for the two- and three-component Gaussian fits, respectively.
    \item $k_1$ and $k_2$ are the number of parameters in the two- and three-component Gaussians, respectively.
    \item $df_2 = N - k_2$ is the degrees of freedom for the more complex fit, where $N$ is the number of data points.
\end{itemize}
The corresponding p-value is obtained from the F-distribution with $(k_2 - k_1)$ numerator and $df_2$ denominator degrees of freedom:
\begin{equation}
    p = 1 - F_{\rm cdf}(F, k_2 - k_1, df_2)
\end{equation}
where $F_{\rm cdf}$ is the cumulative distribution function of the F-distribution.
We  used \texttt{scipy.stats} to calculate the p-value for our models. If $p < 0.05$, the more complex model provides a statistically significant improvement over the simpler model, justifying the need for another Gaussian component. Emission lines where no broad and narrow components are found were excluded from the test. For N103B, the F-test on H$\alpha$ was performed between fitting a Gaussian with five components and four components, to find the intermediate broad component taking the nitrogen  lines into account.

\subsection{Physical parameter determination}
The \texttt{curve\_fit} function from \texttt{scipy} returns the model parameters after performing a successful fit to the emission lines. These parameters include the amplitude $A$, central wavelength $\lambda_0$, and standard deviation $\sigma$. Further mathematical operations are performed with the returned parameters to find physical parameters like surface brightness, velocity width, centroid shift and intensity, necessary for the diagnosis of the Balmer lines. 

The surface brightness is calculated by dividing the total integral of the Gaussian line by the spatial area covered by the ROIs. Each pixel has an area of $0.2\, \textrm{arcsec} \times 0.2\,\textrm{arcsec}$ and the integral is given by:
\begin{equation}
\text{Flux} = \frac{A  \sigma}{0.3989}
\label{eq:gaussian_flux}
\end{equation}
The broad and narrow velocity width is given by:
\begin{equation}
\text{FWHM}_v = 2\sqrt{2\ln{2}} \, \sigma_v \approx 2.3548 \frac{\sigma_\lambda}{\lambda_0}  c
\label{eq:fwhm_velocity}
\end{equation}
where $\sigma_\lambda$ is the Gaussian standard deviation in \AA, $\mathrm{\lambda_0}$ is the central wavelength of the emission line and $c$ is in km\,s$^{-1}$. The Doppler shift velocity is calculated from the difference between the observed central wavelength and the rest wavelength of the line. We calculated the Doppler shift for both the broad and narrow components and the difference between their shifts gives the centroid shift parameter in our analysis.

\subsection{Error propagation}
Errors were propagated using the square root of the diagonal components of the covariance matrix returned by the \texttt{curve\_fit} function after fitting Gaussian lines to the signal. It is given by:
\begin{equation}
\sigma_F^2 \approx \left(\frac{\partial F}{\partial x}\right)^2 \sigma_x^2 + \left(\frac{\partial F}{\partial y}\right)^2 \sigma_y^2 + \dots
\label{eq:error_propagation}
\end{equation}
The above equation calculates the variance  $\sigma_F^2$ of $F$ based on the variances ($\sigma_x^2,\sigma_y^2,...$) of its input variables and their partial derivatives. The datacube contains three header files. The primary header consists of the information about the observation, the second header consists of the data for the observed flux and the third header includes the flux variance. Thus, when the datacube is processed, the flux variances are operated on accordingly and are later extracted to plot the error-bars for the corresponding observed fluxes.
Errors for the physical parameters are calculated from the covariance matrix using standard error propagation methods. The equation used to propagate the error for Doppler velocity is given by:
\begin{equation}
\sigma_{v} = c \, \frac{\sigma_{\lambda}}{\lambda_0}
\label{eq:velocity_dispersion}
\end{equation}
 Doppler velocity errors for both the broad and narrow components were calculated using the above equation and propagated to the error in centroid shift by $\sigma_{\rm centroid}=\sigma_{v\_n}+\sigma_{v\_b}$. Similarly, the formula for the error in the velocity width is given by:
\begin{equation}
\sigma_{w} = 2.35482\,c\,  \sqrt{\left(\frac{\sigma}{\sigma_{\sigma}}\right)^2 + \left(\frac{\lambda}{\sigma_{\lambda}}\right)^2 - 2\, \frac{\sigma \lambda \, \text{pcov}[\sigma, \lambda]}{\sigma_{\sigma} \sigma_{\lambda}}}
\label{eq:velocity_width}
\end{equation}
The $\text{pcov}[\sigma,\lambda$] is given by the covariance matrix and the above equation is used to compute errors for the velocity width in the broad, narrow and intermediate (where applicable) lines.

Equation~\ref{eq:error_propagation} shows how the uncertainty in intensity was calculated. This was used to calculate errors for both the broad and narrow components and was later used to propagate uncertainties in the intensity ratio. Any lines where the
uncertainties of the surface brightness or intensity were consistent with zero were discounted.
It is important to note that this was not done for large errors of the centroid shift, as this
simply implies there is uncertainty in the positioning of the broad component centre, not
necessarily its existence.

\begin{equation}
\sigma_{I} = I \,\sqrt{\left(\frac{A}{\sigma_{A}}\right)^2 + \left(\frac{\sigma}{\sigma_{\sigma}}\right)^2 + 2 \frac{A \sigma \, \text{pcov}[A, \sigma]}{\sigma_{A} \sigma_{\sigma}}}
\label{eq:error_propagation}
\end{equation}

\begin{table}
\caption{\label{table5} Intermediate component parameters for H lines in N103B (Region A and Region B) and SNR~0509 (Region A). $I_i$ is the intensity of the intermediate component normalized to the narrow H$\beta$ component ($I_{H\beta,n}=100$). $W_i$ is the full width at half maximum (FWHM) of the intermediate Gaussian component and $V_i$ is the Doppler velocity (km\,s$^{-1}$), not corrected for Earth's motion.}

\begin{tabular}{lcccccc}
\hline
Region & Line & $I_i$ & $W_i$ [km\,s$^{-1}$] & $V_i$ [km\,s$^{-1}$] \\
\hline
N103B - Region A & H$\beta$ & $7.2\pm1.9$ & $435\pm80$ & $260\pm10$ \\
N103B - Region A & H$\alpha$ &$21\pm2$ & $380\pm50$ & $380\pm50$ \\
N103B - Region B & H$\beta$ & $14\pm2$ & $450\pm30$ & $225\pm15$ \\
N103B - Region B & H$\alpha$ & $20\pm3$ & $429\pm8$ & $300\pm10$ \\
N103B - Region C & H$\beta$ & $12\pm3$ & $450\pm20$ & $273\pm13$ \\
N103B - Region C & H$\alpha$ & $24\pm3$ & $360\pm80$ & $335\pm15$ \\
SNR 0509 - Region A & H$\beta$ & $3.0\pm2.3$ & $290\pm70$ & $300\pm30$ \\
SNR 0509 - Region A & H$\alpha$ & $12.8\pm1.1$ & $296\pm17$ & $270\pm60$ \\
\hline
\end{tabular}

\end{table}

\end{document}